\documentclass[preprintnumbers,superscriptaddress,nofootinbib]{revtex4-2}
\usepackage{hyperref}
\hypersetup{colorlinks=true,linkcolor=purple,anchorcolor=blue,citecolor=blue, filecolor=blue,urlcolor=red,bookmarksnumbered=true,
	pdfview=FitB
}
\usepackage{color}
\usepackage{booktabs} 
\usepackage{threeparttable} 
\usepackage{xcolor}
\usepackage{ulem}
\usepackage{csquotes}
\colorlet{purple1}{blue!70!red}
\colorlet{darkred}{red!50!black}
\usepackage{slashed}
\usepackage{graphicx}
\usepackage[bf,SL,BF]{subfigure}
\usepackage{psfrag}
\usepackage{color}
\usepackage{amssymb}
\usepackage{amsmath}
\usepackage{epstopdf}
\usepackage{natbib}
\usepackage[mathlines]{lineno}

\def\be{\begin{equation}}
\def\ee{\end{equation}}
\def\bea{\begin{eqnarray}}
\def\eea{\end{eqnarray}}
\begin{document}
	
\title{Strong decays of newly observed charm-strange meson states \\ based on a relativistic model}

\author{Bhoomika Pandya}
\email{bhumispandya@gmail.com}
\affiliation{Department of Physics Education, Teachers College, Kyungpook National University, Daegu 41566, Republic of Korea}
\affiliation{Department of Physics, Education Research Center for Quantum Nature of Particles and Matter, Daegu 41566, Republic of Korea }
\author{Manan N Shah}
\email{mnshah09@gmail.com}

\author{P C Vinodkumar}
\email{p.c.vinodkumar@gmail.com}

\affiliation{P. D. Patel Institute of Applied Sciences, Charotar University of Science and Technology, CHARUSAT Campus, Changa- 388421, INDIA}

\begin{abstract}
The mass spectra of charm-strange mesons including ground, radial and orbital excitations have been calculated within a relativistic Dirac framework. The predicted masses of the ground state $1S$ and the first orbital excited state $1P$ exhibits excellent agreement with values reported by the PDG. Utilizing these mass predictions and an effective Lagrangian approach based on heavy quark and chiral symmetries, we have computed the OZI allowed two body strong decay widths of higher excited state candidates of charm-strange sector. The resulting partial widths and branching ratios enable the assignment of spin-parity quantum numbers to several newly observed charm-strange states. In particular, we identify $D_{sJ}(2700)$ as the $2^3S_1$, $D_{s0}(2590)$ as the $2^1S_0$, $D_{sJ}(2860)$ as the $1^3D_3$, $D_{s1}(2860)$ as the $1^3D_1$, and $D_{sJ}(3040)$ as the $2^1P_1$ states. The effective coupling constants $g_T$, $g_X$, $g_Y$, $\tilde{g}_S$ and $\tilde{g}_H$ extracted from present analysis. Furthermore, the $D^*K$ decay channel emerges as the most promising mode for the experimental search of the missing $1^3D_2$ and $1^1D_2$ states, while the $2^3P_0$ state is more likely to be observed through the $DK$ channel.
\end{abstract}

\maketitle


\section{Introduction}\label{intro} 
The renewed interest in the charm strange meson sector is driven by the remarkable discoveries of many new resonances reported by the continuous efforts of the worldwide experimental facilities. The previously existing spectrum of $D_{sJ}$ mesons is enriched by observation of two $D_s^+$ states denoted as $D_{SJ}(2860)^+$ and $X(2690)^+$ by $BABAR$ collaboration in $D^0K^+$ and $D^+K_s^0$ channels, since long. These states were recorded by $BABAR$ detector at 10.6 GeV center of mass energy and sampling of $240 fb^{-1}$ at the $PEP-II$ asymmetric energy $e^+e^-$ storage rings at $SLAC$.  The higher mass structure $D_{SJ}(2860)$ with a narrow width and $X(2690)^+$ with broader width were observed in three independent channels  \cite{BABAR2006DS}. 

The another $B$ factory $Belle$ reported the observation of $D_{sJ}(2700)^+$ with $J^P= 1^-$; in the Dalitz plot analysis of process $B^+ \rightarrow \bar{D}^0 D^0 K^+$ with statistical significance of $8.4 \sigma$. The results were obtained using the 414 $fb^{-1}$ integrated luminosity data along with $449 \times 10^6$ $B\bar{B}$ pairs collected from $Belle$ detector at KEBK \cite{belle2008ds}. The mass range of $D_{sJ}(2700)^+$ was close to previously reported $X(2690)^+$ by $BABAR$ \cite{BABAR2006DS}. Therefore, both of them considered as the same states. The $D_{sJ}(2700)$ and $D_{sJ}(2860)$ states were again confirmed by $BABAR$ in $D^*K$ channel, in 2009 \cite{BABAR2009ds}. This measurement was obtained with $BABAR$ detector at $PEP-II$ asymmetric energy $e^+e^-$ storage rings at $SLAC$ and 470 $fb^{-1}$ sampling. In the case of $D_{sJ}(2700)$, the modification in mass and width was approximately 22 MeV and 37 MeV; respectively than the previous attempt of $BABAR$ \cite{BABAR2006DS}. However the difference found to be negligible for the state $D_{sJ}(2860)$. The same study also reported the measurements of branching ratios $\frac{D^*K}{DK}$ for $D_{sJ}(2700)$ and $D_{sJ}(2860)$ states. Additionally, a new broad structure designated as $D_{sJ}(3040)^+$ also observed in the $D^*K$ mass distribution \cite{BABAR2009ds}. Subsequently, states $D_{s1}^*(2700)^+$ and $D_{sJ}^*(2860)$ were confirmed by the $LHCb$ group and their results were consistent with earlier observations of $BABAR$ and $Belle$ \cite{lhcb2012ds}. These states were found in $D^+K_s^0$ and $D^0K^+$ final states by $LHCb$ detector using $1.0 fb^{-1}$ data in $pp$ collision at $\sqrt{s}= 7$ TeV with improved statistical uncertainties in mass and width as compared to $BABAR$ \cite{lhcb2012ds}. In 2014, the first Dalitz plot analysis of the decay $B_s^0 \rightarrow \bar{D}^0 K^0 \pi^+$ was carried out by $LHCb$ collaboration with configuration of 3.0 $fb^{-1}$ integrated luminosity of $pp$ collision at $\sqrt{s}=$ 7 to 8 TeV \cite{lhcb2014dsprl}. The amplitude analysis suggests one state with spin-1 and another with spin-3 around mass 2860 MeV denoted as $D_{s1}^*(2860)$ and $D_{s3}^*(2860)$ respectively. Though, the mass of spin-1 and spin-3 was found to be same, the width of spin-1 state was three times higher than spin-3 state. Both resonances were believed to be suitable candidates of $1D$ states \cite{lhcb2014dsprl}. More recently, $LHCb$ collaboration studied $B^0 \rightarrow D^-D^+K^+\pi^-$ decay process with an integrated luminosity of 5.4 $fb^{-1}$ of $pp$ collision at center of mass energy of 13 TeV, a new resonance $D_s^+(2590)^+$ decaying into the $D^+K^+\pi^-$ final state was reported. The amplitude analysis suggests $0^-$ parity for $D_s^+(2590)^+$ \cite{lhcb2021ds}. A summary of the detailed experimental data from $BABAR$, $Belle$ and $LHCb$ on charm strange mesons is shown in Table \ref{dsstatus}.

More advances in the future experimental facilities are  expected to further unraveled much information on open flavour mesons. Despite the observation of many states in charm-strange mesons, their spin-parity assignments are still elusive. {{For instance, the nature of the $D_{sJ}(3040)$ meson remains ambiguous with various interpretations proposed. It has been commonly identified as the first radial excitation of the $D_{s1}(2460)$ in Refs.~\cite{Hao2022,chen2009,Sun2009} whereas Ref. \cite{Ebert2010} suggests that the state corresponds instead to the radial excitation of the $D_{s1}(2536)$.
Some analyses have also explored the possibility of interpreting $D_{sJ}(3040)$ as a higher radial excitation $n=3$ or 4 with $J^P = 1^+$ \cite{Segovia2012,Segovia2013}  and as a mixed $1^3D_2$–$1^1D_2$ charmed-strange state \cite{Colangelo2010}. Moreover, exotic interpretations have also been considered, including its possible description as a $D^*(2600)K$ molecular bound state \cite{Guo2011}.
}} 

Due to the presence of single heavy flavour quark in heavy-light mesons, QCD exhibits global symmetry corresponding to the infinite heavy quark mass $(m_Q \rightarrow \infty)$ as compared to the typical QCD scale $\Lambda_{QCD} ( \sim 1 $ GeV$)$. Under such considerations the relative motion of the light flavour quark/antiquark is greatly influenced by the static field provided by the  heavy antiquark/quark. These relativistic effects of light degrees of freedom have a substantial impact on the properties of heavy-light mesons. In this respect, it is more convenient to describe them in the context of Dirac relativistic framework. Following that, we take into account the relativistic Dirac equation with evenly mixed four vector plus scalar power law potential to obtain the single particle confinement energies of quark and antiquark. {{{Previously, two of us investigated the charm-strange meson system within a similar theoretical framework, employing the Martin potential with a confinement term exhibiting $r^{0.1}$ dependence \cite{shah20141}. In this work, we adopt a linear potential, which aligns more closely with the long-distance behavior anticipated by QCD and supported by lattice simulations. This shift is particularly significant for higher excited charm-strange mesons, where the confining dynamics critically influence mass spectra and decay patterns. The use of a linear potential thus provides a more QCD consistent framework for extracting phenomenologically relevant observables and identifying excited states.
}}} Within this framework, the meson masses are computed employing these single particle confinement energies in addition to the spin-spin coupling. We fix the potential parameters for the known ground state and predict the masses of the radial and orbital excited states.

The spin-parity identifications of  resonances can be scrutinized through their various strong decay channels. The concurrent investigations of the mass and strong decay modes provides insights into the internal structure and dynamics of these states. To support such analyses, few models have been developed to address the strong decays of hadrons. The most influential approaches are the pair creation models and the heavy quark effective theory framework. The usual way to deal with the pair creation approach includes the string-breaking model \cite{Alcock1984}, the flux-tube model \cite{Kumano1988,chen2009} and $^3P_0$ model \cite{Sun2009,li2010,luo2009}. Several recent papers have incorporated the $^3P_0$ model to compute the strong decay widths of these newly observed states \cite{Yu2020,Xie2021,Gao2022,Hao2022,Jiang2024}. 
{{
However, a major source of uncertainty in the pair creation models stems from its strong dependence on hadronic wavefunctions, typically modeled using simple harmonic oscillator forms. The oscillator parameter $\beta$ governs the radial and angular structure of these wavefunctions and even small changes in $\beta$ can significantly affect decay amplitudes and branching ratios leading to notable sensitivity in predictions.
Additionally, the pair creation strength $\gamma$, which characterizes vacuum $q\bar{q}$ production is often varies with quark flavor and kinematics, further limiting its consistency and universality across different flavour sector decay processes. A detailed analysis of these limitations is provided in Refs. \cite{Blundell1995,Blundell19951,Yu2014}.}}

{{Within the framework of heavy quark symmetry and chiral effective field theory, collectively known as heavy quark effective theory (HQET) the leading-order hadronic couplings governing strong decays are expected to be  invariant under heavy quark spin and flavor transformations. Also, the allowed range of hadronic coupling constants to lie between 0 and 1. This spin-flavor symmetry implies that coupling constants can be treated as universal parameters across heavy-light meson systems like $D$, $D_s$, $B$, and $B_s$. Consequently, values extracted from the one sector where experimental data are relatively abundant can be reliably extrapolated to the another sector where experimental data is scarce. This universality is particularly valuable for constraining theoretical models to make more consistent predictions of properties of heavy-light mesons. Unlike conventional HQET analyses that rely on experimentally measured masses, the present work adopts a self-contained framework that integrates mass predictions from the Dirac relativistic model with decay dynamics governed by chiral HQET. We believe that this unified approach not only enhances the predictive capability, especially for  identifying higher excited states where experimental data are scarce and theoretical ambiguities persist but also provides strong motivation to revisit and further investigate the charm-strange meson sector. Hence, in the present work we use Dirac formalism to predict the masses of hadrons. These predicted masses are then used as inputs to compute decay amplitudes for Okubo-Zweig-Iizuka (OZI) allowed channels using effective field theory. Where necessary, previously determined coupling constants are employed to make further predictions. Once the experimental identification of a state is confirmed, we extract the relevant coupling constants and utilize them to predict the properties of spin partner states as searching for the spin partner of higher excited states remains important objective of ongoing research.
}}
 
We begin in Sec. \ref{TF} with a description of the relativistic Dirac framework for heavy and light quarks. The computational details of obtaining mass spectra of the heavy-light flavour mesons are also discussed in this section. In Sec. \ref{strongdecays} the strong decays of heavy-light mesons within the heavy quark effective theory formalism proposed by P. Colangelo is outlined.  The numerical results on mass and strong decay calculations are presented in Sec. \ref{RD}, followed by discussions based on our analysis to identify the spin-parity for many of the new states in the charm-strange sector. In the end, a summary of the present study is presented in Sec. \ref{summary}.

\begin{table*}
\begin{center}
\tabcolsep 15pt
\caption{Experimental information of new charm- strange resonances observed by $BABAR$, $Belle$ and $LHCb$ collaboration.}\label{dsstatus}
\begin{tabular}{lccccc}
\hline\hline\noalign{\smallskip}
Resonance & Mass (in MeV) & Width (in MeV) & $J^P$ & Observation & Reference \\
\noalign{\smallskip}\hline\noalign{\smallskip}
$D_{s1}^*(2700)$    & 2688 $\pm$ 4 $\pm$ 4 & 112$\pm$7$\pm$36 &  & $BABAR$ & \cite{BABAR2006DS}      \\
$D_{sJ}^*(2860)$    & 2856.6 $\pm$ 1.5 $\pm$ 5.0 & 47$\pm$7$\pm$10 &  & $BABAR$ & \cite{BABAR2006DS} \\
$D_{s1}^*(2700)$    & $2708 \pm 9 ^{+11}_{-10}$  & $108\pm23^{+36}_{-31}$ & $1^{-}$ & $Belle$   & \cite{belle2008ds}    \\
$D_{s1}^*(2700)$    & $2710 \pm 2 ^{+12}_{-7}$  & $149\pm7^{+39}_{-52}$ &  & $BABAR$ & \cite{BABAR2009ds}      \\
$D_{sJ}^*(2860)$    & $2862 \pm 2 ^{+5}_{-2}$ & 48$\pm$3$\pm$6 &  & $BABAR$ & \cite{BABAR2009ds} \\
$D_{sJ}^*(3040)$    & $3044 \pm 8 ^{+30}_{-5}$ & $239\pm35^{+46}_{-42}$ &  & $BABAR$ & \cite{BABAR2009ds}\\
$D_{s1}^*(2700)$    &  $2709.2 \pm 1.9 \pm 4.5$ & $115.8 \pm 7.3 \pm 12.1$   &  & $LHCb$  & \cite{lhcb2012ds}     \\
$D_{sJ}^*(2860)$    &  $2866.1 \pm 1.0 \pm 6.3$ & $69.9 \pm 3.2 \pm 6.6$   &  & $LHCb$ & \cite{lhcb2012ds}      \\
$D_{s3}^*(2860)$    &  $2860.5 \pm 2.6 \pm 2.5 \pm 6.0$ & $53 \pm 7 \pm 4 \pm 6$   & $3^{-}$ & $LHCb$ & \cite{lhcb2014dsprl}      \\
$D_{s1}^*(2860)$    &  $2859 \pm 12 \pm 6 \pm 23$ & $159 \pm 23 \pm 27 \pm 72$   & $1^{-}$ & $LHCb$  & \cite{lhcb2014dsprl}     \\
$D_{s0}^*(2590)$    &  $2591 \pm 6 \pm 7 $ & $89 \pm 16 \pm 12$   & $0^{-}$ & $LHCb$ & \cite{lhcb2021ds}      \\
\noalign{\smallskip}\hline\hline
\end{tabular}
\end{center}
\end{table*} 
\section{Theoretical Framework}\label{TF}
\subsection{ Confined energies of the quarks and anti-quarks based on the relativistic model}
{{Heavy-light mesons, composed of a heavy quark $Q$ and a light antiquark $\bar{q}$, are systems where relativistic effects play a crucial role due to the significant mass difference between the constituents. These effects are naturally incorporated within the relativistic Dirac formalism, which provides a consistent framework for describing their structure and dynamics.}} We employ the central confinement potential for quark and anti-quark as Lorentz scalar plus vector linear term given by \cite{Barik1985,pandya2021, Shah2016bbs,Bhavsar2018} 
\begin{equation}\label{eq:a}
V(r)= \frac{1}{2} (1+\gamma_0) (\lambda r^{\nu}+V_0)
\end{equation}
where $\nu =1.0$ and $r$ is the radial distance from the meson centre of mass. Then  strength of the potential $({\lambda})$ and depth of the potential $(V_0)$ are  phenomenological parameters left in the formalism to adjust from the known ground state.  \\
The constituents quark and anti-quark field can be expressed through the Lagrangian density as \cite{Barik1985}
\begin{equation}\label{eq:b}
{\mathcal L} (r)= \bar{\psi}(r) \left[\frac{i}{2} \gamma^{\mu} \overrightarrow{\partial_\mu} - V(r) - m_q \right] \psi(r)
\end{equation}
Then the independent quark wave function $\psi(\vec{r})$ satisfies the Dirac equation given as \cite{Barik1985, Bhavsar2018}
\begin{equation}\label{eq:c}
[\gamma^0 E - \vec{\gamma}. \vec{P} - m_q - V (r)]\psi (\vec{r}) = 0.
\end{equation}
where $E$ is the individual quark confined energy. 

\begin{table}
\begin{center}
\tabcolsep 8pt
\caption{The quark masses and potential parameters for $D_s$ mass spectrum.}\label{input}
\begin{tabular}{llr}
\hline\hline
& Parameter & Value  \\
\hline
Quark Mass & $m_c$ (GeV) & 1.27  \\
           & $m_s$ (GeV) & 0.093  \\
Potential Depth & $V_0$  (GeV)    & -0.0001 \\
Potential Strength & $\lambda$ (GeV$^{2}$ ) & $0.0405 (2n + l+ 1)^{0.5}$  \\

$jj$ coupling strength  & $\sigma$  (GeV$^{3}$) & 0.159  \\
\hline\hline
\end{tabular}
\end{center}
\end{table}

\begin{table*}
\begin{center}
\tabcolsep 7pt
\small
\caption{S-wave $D_s$ meson spectrum (in MeV).} \label{swds}
\begin{tabular}{cccclll}
\hline\hline
&       &&    & \multicolumn{3}{c} {Experiment} \\
\cline{5-7} \smallskip
State $(s_lJ^P)$ & $M_{Q \bar {q}}$ & $\langle V_{j_1j_2}\rangle$ & Present  &   PDG \cite{PDG2020} & $BABAR$ \cite{BABAR2006DS,BABAR2009ds} & $LHCb$ \cite{lhcb2012ds,lhcb2021ds}\\
\hline
$1{^3S_1} (_\frac{1}{2}1^-)$ & 2102.21 & 10.92 & 2113.14 & $D_s^*$  $(2112.2 \pm 0.4)$ & \ldots & \ldots\\
$1{^1S_0} (_\frac{1}{2}0^-)$ & 2001.74 & -32.78 & 1968.95 & $D_s$ (1867.83 $\pm$ 0.05) & \ldots & \ldots \\
$2{^3S_1} (_\frac{1}{2}1^-)$ & 2725.05 & 6.46 & 2731.51 &  $?D_{s1}^*(2700)  (2714 \pm 5)$ & $?D_{s1}^*(2700)  (2688 \pm 4)$ & $?D_{s1}^*(2700)  (2709.2 \pm 1.9)$  \\
 &  &  &  & \ldots  & $?D_{s1}^*(2700)(2710 \pm 2)$ &  \ldots \\
$2{^1S_0} (_\frac{1}{2}0^-)$ & 2629.11 & -19.38 & 2609.73 & \ldots & ?$D_{sJ}^*(3040)$ $(3044 \pm 8)$ & \ldots  \\
 &  & &  & \ldots & \ldots & ?$D_{s0}(2590)$ $(2591 \pm 6)$   \\
$3{^3S_1} (_\frac{1}{2}1^-)$ & 3337.02 & 4.44  & 3341.46 & \ldots &   \ldots & \ldots    \\
$3{^1S_0}(_\frac{1}{2}0^-)$ & 3244.32 & -13.33  & 3230.99 &  \ldots          & \ldots  &    \ldots   \\
$4{^3S_1}$ $(_\frac{1}{2}1^-)$ & 3875.32 & 3.40 & 3878.72 &           $\cdots$  & \ldots & \ldots       \\
$4{^1S_0}$ $(_\frac{1}{2}0^-)$ & 3787.26 & -10.20  & 3777.05 &           $\cdots$   & \ldots & \ldots       \\
$5{^3S_1}$ $(_\frac{1}{2}1^-)$ & 4224.96 & 2.91 & 4227.88 &           $\cdots$ & \ldots & \ldots        \\
$5{^1S_0}$ $(_\frac{1}{2}0^-)$ & 4145.00 & -8.75  & 4136.24 &           $\cdots$  & \ldots & \ldots       \\
\hline\hline\smallskip
\end{tabular}
\end{center}
\end{table*}

The corresponding two component solution to the normalised independent quark wave function denoted as \cite{Greiner2000,pandya2021, Bhavsar2018,shah20141}
\begin{equation}\label{eq:d}
\psi_{nlj}(r) = \left(
    \begin{array}{c}
      \psi_{nlj}^{(+)} \\
      \psi_{nlj}^{(-)}
    \end{array}
  \right)
\end{equation}
where
\begin{equation}\label{eq:e}
\psi_{nlj}^{(+)}(\vec{r}) = N_{nlj} \left(
    \begin{array}{c}
      i g(r)/r \\
      (\sigma.\hat{r}) f(r)/r
    \end{array}
  \right) {\cal{Y}}_{ljm}(\hat{r})
\end{equation}
\begin{equation}\label{eq:f}
\psi_{nlj}^{(-)}(\vec{r}) = N_{nlj} \left(
    \begin{array}{c}
      i (\sigma.\hat{r}) f(r)/r \\
        g(r)/r
    \end{array}
  \right) (-1)^{j+m_j-l} {\cal{Y}}_{ljm}(\hat{r})
\end{equation}\\
and $N_{nlj}$ is the overall normalization constant \cite{Barik1983, Greiner2000, Shah2016bbs, Bhavsar2018}. Here, the normalized spin angular part is defined by \cite{Greiner2000}
\begin{equation}\label{eq:g}
{\cal{Y}}_{ljm}(\hat{r}) = \sum_{m_l, m_s}\langle l, m_l, \frac{1}{2}, m_s| j, m_j \rangle \, Y^{m_l}_l \chi^{m_s}_{\frac{1}{2}} 
\end{equation}
The two spinors  $\chi_{\frac{1}{2}{m_s}}$ are the eignefunctions of the spin operators and explicitly written as \cite{Greiner2000}
\begin{equation}\label{eq:h}
\chi_{\frac{1}{2} \frac{1}{2}} = \left(
    \begin{array}{c}
      1 \\
      0
    \end{array}
  \right) \ \ \ , \ \ \ \ \chi_{\frac{1}{2} -\frac{1}{2}} = \left(
    \begin{array}{c}
      0 \\
      1
    \end{array}
  \right)
\end{equation}
The reduced radial parts $g(r)$ and $f(r)$ then satisfy the following equations \cite{pandya2021,Bhavsar2018,shah20141, Shah2016bbs}
\begin{equation}\label{eq:i1}
\frac{d^2 g(r)}{dr^2}+\left[(E + m_q) [E - m_q - V(r)] - \frac{\kappa (\kappa + 1)}{r^2}\right] g(r) = 0
\end{equation}
and
\begin{equation}\label{eq:i2}
\frac{d^2 f(r)}{dr^2}+\left[(E + m_q) [E - m_q - V(r)] - \frac{\kappa (\kappa - 1)}{r^2}\right] f(r) = 0
\end{equation}
with the defined of quantum number $\kappa $ as \cite{Greiner2000}
\begin{eqnarray}\label{eq:kappa}
\kappa = &\left\{\begin{matrix} -(\ell + 1)& = - \left(j+\frac{1}{2}\right) & \ \ for \ \ j= \ell+\frac{1}{2}\\
                                    \ell & = + \left(j+\frac{1}{2}\right) & \ \ for \ \ j= \ell-\frac{1}{2} \end{matrix}\right.
\end{eqnarray}
Now taking the form of $V(r)$ as given in Eq. (\ref{eq:a}) and introducing a dimensionless variable $\rho=\frac{r}{r_0}$ with the conveniently chosen scale factor as \cite{Jena1983, Barik1983, Bhavsar2018, Shah2016bbs}
\begin{equation}\label{eq:k}
r_0 = \left[(m_q + E)\frac{\lambda}{2}\right]^{-\frac{1}{3}},
\end{equation}
reduces the equations (9) and (10)  to the  Schr\"{o}dinger like form \cite{Jena1983, Jena19831, Barik1983, Bhavsar2018, Shah2016bbs}
\begin{equation}\label{eq:j1}
\frac{d^2 g(\rho)}{d\rho^2}+\left[\epsilon- \rho - \frac{\kappa (\kappa+1)}{\rho^2}\right] g(\rho) = 0
\end{equation}
 and
\begin{equation}\label{eq:j2}
\frac{d^2 f(\rho)}{d\rho^2}+\left[\epsilon- \rho - \frac{\kappa (\kappa -1)}{\rho^2}\right] f(\rho) = 0
\end{equation}
where
\begin{equation}\label{eq:l}
\epsilon = (E - m_q - V_0) (m_q + E)^{\frac{1}{3}} \left(\frac{\lambda}{2}\right)^{\frac{-2}{3}}
\end{equation}
is the dimensionless energy eigenvalue related to the confined bound state of quarks \cite{Jena1983, Jena19831, Barik1983, Bhavsar2018, Shah2016bbs}.
Now it is possible to solve expressions (\ref{eq:j1}) and (\ref{eq:j2}) numerically with the proper choice of $\kappa$. 
Also, the solutions $g (\rho)$ and $f (\rho)$ are normalized to get
\begin{equation}
 \int_0^\infty (f_q^2(\rho) + g_q^2(\rho)) \ d \rho = 1.
\end{equation}
\subsection{Masses of the heavy-light mesonic system}
After getting the confined energies of quark and anti-quark one can construct the wavefunctions for heavy-light meson as the symmetric combinations of the positive energy solutions $\psi_{Q}$ and the negative energy solutions $\psi_{\bar{q}}$ of Eq.(\ref{eq:e}) and Eq.(\ref{eq:f}). The corresponding confinement energy of the $Q\bar{q}$ system can be written as
\begin{equation}
M_{Q \bar{q}} = E_{Q} + E_{\bar{q}}
\end{equation}
where $E_{Q/\bar{q}}$ are obtained from expression (\ref{eq:l}) which also include the centrifugal repulsion of the center of mass. 

\begin{table*}
\begin{center}
\tabcolsep 1.5pt
\small
\caption{P-wave $D_s$ meson spectrum (in MeV).} \label{pwds}
\begin{tabular}{cccccclll}
\hline\hline
 &   &&&&           & \multicolumn{3}{c} {Experiment} \\
\cline{7-9} 
 State $(s_lJ^P)$ & $M_{Q \bar{q}}$ & $\langle V_{j1j2}\rangle$ & $\langle V_{LS}\rangle$ & $\langle V_{T}\rangle$ & Present  &  PDG \cite{PDG2020} & $BABAR$ \cite{BABAR2006DS,BABAR2009ds} & $LHCb$ \cite{lhcb2012ds}  \\
\hline\smallskip
$1{^3P_2}$ $(_\frac{3}{2}2^+)$  & 2560.08 & -10.99  & 13.85  & -1.53 & 2561.41 & $D_{s2}^*(2573)(2569.1 \pm 0.8)$ &  \ldots & $\ldots$ \\
$1{^3P_1}$  $(_\frac{3}{2}1^+)$ & 2560.08 & -36.65 & -13.85 & 7.67 & 2517.25 & $D_{s1}(2536)(2535.11 \pm 0.06)$ & \ldots & \ldots \\
$1{^3P_0}$  $(_\frac{1}{2}0^+)$ & 2560.08 & -21.98  & -27.71 & -15.34 & 2454.67 & $D_{s0}(2317)(2317.8 \pm 0.8)$ & \ldots & \ldots  \\
$1{^1P_1}$ $(_\frac{1}{2}1^+)$ & 2452.06 & 86.55  & 0 & 0 & 2538.62 & $D_{1s}(2460)(2459.5 \pm 0.6)$ & \ldots & \ldots \\

$2{^3P_2}$  $(_\frac{3}{2}2^+)$ & 3065.64 & -7.76 & 10.52  & -1.16   & 3067.23  &   $\ldots$  & $?D_{sJ}^*(2860)(2856.6 \pm 1.5)$ & $?D_{sJ}^*(2860)(2866.1 \pm 1.0)$   \\
&   &   &  &  &  & \ldots  & $?D_{sJ}^*(2860)(2862 \pm 2)$ & \ldots   \\
$2{^3P_1}$  $(_\frac{3}{2}1^+)$ &  3065.64 & -25.88 & -10.52  & 5.82 & 3035.06 &  \ldots & ?$D_{sJ}^*(3040)(3044 \pm 8)$ & \ldots   \\
$2{^3P_0}$  $(_\frac{1}{2}0^+)$&  3065.64 & -15.53  & -21.04 & -11.65 & 3017.41 &   $\ldots$  & $?D_{sJ}^*(2860)(2856.6 \pm 1.5)$ & $?D_{sJ}^*(2860)$ $(2866.1 \pm 1.0)$    \\
&   &   &  &  &  &  \ldots & $?D_{sJ}^*(2860)(2862 \pm 2)$ &  \ldots  \\
$2{^1P_1}$  $(_\frac{1}{2}1^+)$ & 2969.3 & 57.79 &     0   &  0   & 3027.09 & \ldots & ?$D_{sJ}^*(3040)(3044 \pm 8)$ & \ldots\\
$3{^3P_2}$  $(_\frac{3}{2}2^+)$ & 3559.57 & -5.92 & 8.36 & -0.93 & 3561.08 & \ldots & \ldots & \ldots \\
$3{^3P_1}$  $(_\frac{3}{2}1^+)$ &  3559.57 & -19.75 & -8.36 & 4.63 & 3536.09 & \ldots & \ldots & \ldots\\
$3{^3P_0}$  $(_\frac{1}{2}0^+)$ &  3559.57 & -11.85 & -16.73 & -9.27 & 3521.72  & \ldots & \ldots & \ldots\\
$3{^1P_1}$  $(_\frac{1}{2}1^+)$ & 3470.33 & 42.88  &  0  & 0      & 3513.22 & \ldots & \ldots & \ldots\\
$4{^3P_2}$  $(_\frac{3}{2}2^+)$ & 4042.25 & -4.73 & 7.37 & -0.82 & 4044.1  & \ldots & \ldots & \ldots \\
$4{^3P_1}$  $(_\frac{3}{2}1^+)$ & 4042.25 & -15.76 & -7.37 & 4.08 & 4023.2  & \ldots & \ldots & \ldots\\
$4{^3P_0}$  $(_\frac{1}{2}0^+)$ & 4042.25 & -9.45 & -14.74 & -8.17 & 4009.9 & \ldots & \ldots & \ldots\\
$4{^1P_1}$  $(_\frac{1}{2}1^+)$ & 3957.79 & 33.65  &  0  & 0      & 3991.4 & \ldots & \ldots & \ldots\\   
$5{^3P_2}$  $(_\frac{3}{2}2^+)$ & 4484.26 & -3.93 & 6.43 & -0.71 & 4486 & \ldots & \ldots & \ldots \\
$5{^3P_1}$  $(_\frac{3}{2}1^+)$ & 4484.26 & -13.13 & -6.43 & 3.56 & 4468.3 & \ldots & \ldots & \ldots\\
$5{^3P_0}$  $(_\frac{1}{2}0^+)$ & 4484.26 & -7.87 & -12.86 & -7.13 & 4456.4 & \ldots & \ldots & \ldots\\
$5{^1P_1}$  $(_\frac{1}{2}1^+)$ & 4404.11 & 27.71  &  0  & 0    & 4431.8 & \ldots & \ldots & \ldots\\  
\hline\hline\smallskip
\end{tabular}
\end{center}
\end{table*}

\begin{table*}[htp]
\begin{center}
\tabcolsep 3.5pt
 \small
\caption{D-wave and F-wave $D_s$ meson spectrum (in MeV).} \label{dfwds}
\begin{tabular}{cccccclll}
\hline\hline
 &   &&&&           & \multicolumn{3}{c} {Experiment} \\
\cline{7-9}
 State $(s_lJ^P)$ & $M_{Q \bar {q}}$ &$\langle V_{j1j2}\rangle$ &$ \langle V_{LS}\rangle$ & $\langle V_{T}\rangle$ & Present  &  PDG \cite{PDG2020} & $BABAR$ \cite{BABAR2006DS,BABAR2009ds} & $LHCb$ \cite{lhcb2012ds,lhcb2014dsprl}  \\
\hline
$1{^3D_3}$  $(_\frac{5}{2}3^-)$ & 2874.93 & -43.32 & 24.17 & -1.91 & 2853.88 & \ldots & $?D_{sJ}^*(2860)(2856.6 \pm 1.5)$ & $?D_{sJ}^*(2860) (2866.1 \pm 1.0)$   \\
&   &   &  &  &  &  \ldots & $?D_{sJ}^*(2860)(2862 \pm 2)$ & \ldots   \\
  &   &   &  &  &  & \ldots  & \ldots & $?D_{s3}^*(2860)(2860.5 \pm 2.6)$   \\
$1{^3D_2}$  $(_\frac{5}{2}2^-)$ & 2874.93 & -38.82 & -12.08 & 6.69 & 2830.71 & \ldots  & ?$D_{sJ}^*(3040)(3044 \pm 8)$ & \ldots \\
$1{^3D_1}$  $(_\frac{3}{2}1^-)$ & 2874.93 & -12.40 & -36.25 & -6.69 & 2819.57 &  $?D_{s1}^*(2700)  (2714 \pm 5)$ & $?D_{s1}^*(2700)(2688 \pm 4)$ & $?D_{s1}^*(2700)(2709.2 \pm 1.9)$  \\
 &  &  &  & &  & \ldots & $?D_{s1}^*(2700)(2710 \pm 2)$ & \ldots \\
&   &   &  &  &  & \ldots  &  $?D_{sJ}^*(2860)(2856.6 \pm 1.5)$  & \ldots   \\ 
 &   &   &  &  &  & \ldots  & $?D_{sJ}^*(2860)(2862 \pm 2)$ & $?D_{sJ}^*(2860)$ $(2866.1 \pm 1.0)$   \\
  &   &   &  &  &  & \ldots  & \ldots  & $?D_{s1}^*(2860)(2859 \pm 12)$   \\
$1{^1D_2}$  $(_\frac{3}{2}2^-)$ & 2772.86 & 75.77  & 0    & 0   & 2848.63 & \ldots & ?$D_{sJ}^*(3040)(3044 \pm 8)$ & \ldots \\


$2{^3D_3}$ $(_\frac{5}{2}3^-)$& 3484.49 & -30.08 & 19.49  & -1.54 & 3472.35 &  \ldots & \ldots & \ldots\\
$2{^3D_2}$ $(_\frac{5}{2}2^-)$& 3484.49 & -26.96 & -9.75 & 5.40 & 3453.18 &    \ldots & \ldots & \ldots     \\
$2{^3D_1}$ $(_\frac{3}{2}1^-)$& 3484.49 & -8.61 & -29.24 & -5.40 & 3441.23 &     \ldots & \ldots & \ldots  \\
$2{^1D_2}$ $(_\frac{3}{2}2^-)$& 3384.9 & 50.85 & 0    & 0   & 3435.75 &    \ldots & \ldots & \ldots   \\

$3{^3D_3}$ $(_\frac{5}{2}3^-)$ & 3972.65 & -23.83  & 16.29 & -1.29 & 3963.82 &     \ldots & \ldots & \ldots     \\
$3{^3D_2}$ $(_\frac{5}{2}2^-)$& 3972.65 & -21.36 & -8.15 & 4.51 & 3947.66 &   \ldots & \ldots & \ldots        \\
$3{^3D_1}$ $(_\frac{3}{2}1^-)$& 3972.65 & -6.82 &-26.45 & -4.51 & 3936.86   &  \ldots & \ldots & \ldots      \\
$3{^1D_2}$ $(_\frac{3}{2}2^-)$ & 3880.05 & 39.41 & 0   & 0     & 3919.47 &    \ldots & \ldots & \ldots     \\
$4{^3D_3}$ $(_\frac{5}{2}3^-)$& 4389.34 & -19.99  & 13.99 & -1.10 & 4382.23 &     \ldots & \ldots & \ldots      \\
$4{^3D_2}$ $(_\frac{5}{2}2^-)$& 4389.34 & -17.91 & -6.99 & 3.88 & 4368.31 &     \ldots & \ldots & \ldots     \\
$4{^3D_1}$ $(_\frac{3}{2}1^-)$& 4389.34 & -5.72 & -20.98 & -3.88 & 4358.75 &    \ldots & \ldots & \ldots       \\
$4{^1D_2}$ $(_\frac{3}{2}2^-)$& 4303.37 & 32.59 & 0   & 0 & 4335.96 &      \ldots & \ldots & \ldots  \\ 
$1{^3F_4}$ $(_\frac{7}{2}4^+)$& 3332.41 & -16.27 & 31.81 & -1.95 & 3346.01 &   \ldots  & \ldots &   \ldots   \\
$1{^3F_3}$  $(_\frac{7}{2}3^+)$& 3332.41 & -24.71 & -10.60 & 5.87 & 3302.97 &  \ldots &  \ldots &  \ldots    \\
$1{^3F_2}$  $(_\frac{5}{2}2^+)$& 3332.41 & -7.83 & -42.42 & -4.70 & 3277.45 &   $\ldots$ & $?D_{sJ}^*(2860)(2856.6 \pm 1.5)$ & $?D_{sJ}^*(2860)$ $(2866.1 \pm 1.0)$   \\
&   &   &  &  &  &  \ldots &  $?D_{sJ}^*(2860)(2862 \pm 2)$ &   \ldots \\
$1{^1F_3}$  $(_\frac{5}{2}3^+)$ & 3224.98 & 55.78 & 0   & 0     & 3280.76 &  \ldots &  \ldots    & \ldots \\ 
\hline\hline
\end{tabular}
\end{center}
\end{table*}

The mass of the state $M_{^{2 S+1}L_J}$ is obtained by adding the contributions from spin-spin, spin-orbit and tensor interactions of the confined one gluon exchange potential (COGEP) between quark and anti-quark \cite{vinodkumar1992,khadkikar1991} to the $M_{Q\bar{q}}$.                                                   
\begin{equation}
M_{^{2 S+1}L_J} =  M_{Q \bar q} \ (n_1l_1j_1, n_2l_2j_2) +  \langle V_{Q \bar q}^{j_1j_2}\rangle  + \langle V_{Q \bar q}^{LS}\rangle + \langle V_{Q \bar q}^{T}\rangle
\end{equation}

The spin-spin coupling  part is defined from COGEP as \cite{vinodkumar1992,khadkikar1991}
\begin{equation}
\langle V^{j_1 j_2}_{Q \bar q} (r)\rangle = \frac{\sigma \ \langle j_1 j_2 J M |\hat{j_1}.\hat{j_2}| j_1 j_2 J M \rangle}{(E_Q + m_{Q})(E_{\bar{q}} + m_{\bar{q}})}
\end{equation}
where $\sigma$ represents the $jj$ coupling constant. The angular brackets appearing in the term $\langle j_1 j_2 J M |\hat{j_1}.\hat{j_2}| j_1 j_2 J M \rangle$ contains square of Clebsch-Gordan coefficients. \\
The tensor and spin-orbit parts of COGEP  have the form \cite{vinodkumar1992,khadkikar1991}
\begin{eqnarray}\label{Vt}
V^{T}_{Q \bar q} (r) &=& - \frac{\alpha_s}{4} \frac{N_Q^2 N_{\bar q}^2}{(E_Q + m_{Q})(E_{\bar{q}} + m_{\bar{q}})} \nonumber \\ && \otimes \ \lambda_Q . \lambda_{\bar q} \left( \left( \frac{D''_1 (r)}{3}- \frac{D'_1 (r)}{3 \ r} \right) S_{Q \bar q}\right)
\end{eqnarray}

\begin{eqnarray}\label{Vls}
V^{LS}_{Q \bar q} (r) &=& \frac{\alpha_s}{4} \frac{N_Q^2 N_{\bar q}^2}{(E_Q + m_{Q})(E_{\bar{q}} + m_{\bar{q}})}  \frac{\lambda_Q . \lambda_{\bar q}}{2 \ r} \\ &&\otimes \left[ \left[ \vec{r} \times (\hat{p_Q}-\hat{p_q}). (\sigma_Q + \sigma_q)\right]\left( {D'_0 (r)}+ 2 {D'_1 (r)} \right) \right. \nonumber\\ &&  \left. + \left[ \vec{r} \times (\hat{p_Q}+\hat{p_q}). (\sigma_i - \sigma_j)\right]\left( {D'_0 (r)}-  {D'_1 (r)} \right) \right]  \nonumber
\end{eqnarray}
where $S_{Q \bar q} = \left[ 3 (\sigma_Q. {\hat{r}})(\sigma_{\bar q}. {\hat{r}})- \sigma_Q . \sigma_{\bar q}\right]$, ${\hat{r}} = {\hat{r}}_Q - {\hat{r}}_{\bar q}$ is the unit vector in the direction of $\vec{r}$ and $\alpha_s$ is the running strong coupling constant expressed as
 \begin{equation}
 \alpha_s = \frac{4 \pi}{(11-\frac{2}{3}\  n_{f})\log\left(\frac{M^2_Q}{\Lambda^2_{QCD}}\right)}
 \end{equation}
$n_f$ is number of flavours and for charmed mesons we take it as 3. The term ${\langle \lambda_{Q}. \lambda_{\bar{q}} \rangle} = -\frac{4}{3}$ is the color factor and independent of the flavour contents of the quarks \cite{Close}.

The confined gluon propagators are given by \cite{vinodkumar1992,khadkikar1991,monteiro}
\begin{equation}
D_0 (r) = \left( \frac{\alpha_1}{r}+\alpha_2 \right) \exp(-r^2 c_0^2/2)
\end{equation}

\begin{equation}
D_1 (r) =  \frac{\gamma}{r} \exp(-r^2 c_1^2/2)
\end{equation}
with $\alpha_1$ = 1.035, $\alpha_2$ = 0.3977 GeV, $c_0$ = 0.3418 GeV, $\gamma$ = 0.8639 and $c_1$ = 0.4123 GeV GeV as in the previous studies \cite{monteiro,pandya2021,Bhavsar2018}. The model parameters used in present study are listed in Table \ref{input}. The numerical results obtained for the mass spectra are presented in Table \ref{swds} \ref{pwds} and \ref{dfwds}. Also our predictions are compared with available experimental data and the corresponding averaged values reported by the Particle Data Group (PDG) \cite{PDG2020}.

\section{Strong decays of heavy-light mesons in HQET}\label{strongdecays}
The hadrons with the single heavy quark systematically described by considering the heavy quark mass limit $m_Q \rightarrow \infty$ in the framework of heavy quark effective theory. Within this framework, the heavy quark spin symmetry and heavy quark flavour symmetry arises
because the chromomagnetic and flavor-changing interactions of the heavy quark become suppressed by powers of $1/m_Q$. As a result, the heavy quark acts as a static color source and its spin and flavor do not influence the dynamics of the light degree of freedom. Consequently, the QCD Lagrangian becomes approximately invariant under independent rotations of the heavy quark spin and under interchange of heavy quark flavors (e.g., $c \leftrightarrow b$) \cite{Colangelo2006}. These symmetries allows to relate the description of hadrons which differ in spin orientation and heavy quark flavour. The heavy-light  $(Q\bar{q})$ mesons are then classified according to the decoupling of the heavy-quark spin $s_Q$ from the total angular momentum $s_l$ of the light degrees of freedom. Both the heavy quark spin and total angular momentum of light d.o.f are separately conserved in strong interactions. Therefore, the heavy meson states come into the doublets as per the different value of $s_l$. The pair of states in each doublet referred as the spin partners having total spin $J = s_l \pm \frac{1}{2}$ and parity $P= (-1)^{l+1}$ where $l$ is the orbital angular momentum of light d.o.f. and $\vec{s}_l = \vec{l} + \vec{s}_q,$ $\vec{s}_q$ being the spin of light antiquark. 

Accordingly, for the lowest lying states $l=0$ ($S$-wave states of the quark model), $s^P_l=\frac{1}{2}^-$ and doublet consists of two states represented by $(P,P^*)$ with spin-parity $J^P = (0^-,1^-)$ . For $l=1$ ($P$-wave states) is associated with either $s^p_l=\frac{1}{2}^+$ or $s^p_l=\frac{3}{2}^+$. Then two doublets represented as $(P^*_0,P^{\prime}_1)$ and $(P_1,P^*_2)$ with spin-parity given as $J^P=(0^+,1^+)$ and $J^P=(1^+,2^+)$ respectively. For higher orbital excited states, $l=2$ ($D$-wave states) therefor $s^p_l=\frac{3}{2}^-$ and $s^p_l=\frac{5}{2}^-$. Here, one doublet is expressed as $(P^*_1,P_2)$ where $J^P=(1^-,2^-)$ and another as 
$(P^{\prime}_2,P^*_3)$ with $J^P=(2^-,3^-)$. Similarly, for $l=3$ ($F$-wave states) one can write  $s^p_l=\frac{5}{2}^+$ and $s^p_l=\frac{7}{2}^+$ and associated with doublets $(P^*_2,P_3)$ and $(P^{\prime}_3,P^*_4)$ which belongs to    
$J^P=(2^+,3^+)$ and $J^P=(3^+,4^+)$, respectively.  
Both spin and flavour symmetry is then implemented conveniently by associating the fields with each doublets. Accordingly, these fields ($4\times4$  matrices) are presented by $H_a$, $S_a$, $T_a$, $X_a$, $Y_a$, $Z_a$ and $R_a$ as \cite{Falk1992}. 
\begin{equation}
H_a = \frac{1+\slashed{v}}{2} \{P^*_{a\mu} \gamma^{\mu} - P_a \gamma_5 \}
\end{equation}
\begin{equation}
S_a = \frac{1+\slashed{\nu}}{2} \{P^\mu_{1a} \gamma_{\mu} \gamma_5 - P^*_{0a}  \}
\end{equation}

\begin{equation}
\small{
T^{\mu}_a = \frac{1+\slashed{\nu}}{2} \bigg \{P^{*\mu\nu}_{2a} \gamma_{\nu} - P_{1a\nu} \sqrt{\frac{3}{2}}\gamma_5 \left[g^{\mu\nu}-\frac{\gamma^{\nu}\left(\gamma^\mu-\nu^\mu\right)}{3}\right]\bigg \}}
\end{equation}

\begin{equation}
\small{
X^{\mu}_a = \frac{1+\slashed{\nu}}{2} \bigg\{P^{\mu\nu}_{2a} {\gamma_5} {\gamma_{\nu}} - P^*_{1a\nu} \sqrt{\frac{3}{2}} \gamma_5 \left[g^{\mu\nu}-\frac{\gamma_{\nu}\left(\gamma^\mu + \nu^\mu\right)}{3}\right]\bigg \}}
\end{equation}


\begin{eqnarray}\label{yfield}
&& Y^{\mu\nu}_a =  \frac{1+\slashed{\nu}}{2} \bigg \{P^{*\mu\nu\sigma}_{3a} \gamma_{\sigma} - P^{\alpha\beta}_{2a} \sqrt{\frac{5}{3}}\gamma_5  \nonumber\\ && \left[g^{\mu}_{\alpha} g^{\nu}_{\beta} -\frac{g^{\nu}_{\beta} \gamma_{\alpha}\left(g^\mu - \nu^\mu\right)}{5} -\frac{g^{\mu}_{\alpha}\gamma_{\beta}\left(\gamma^\nu - \nu^\nu\right)}{5}\right]\bigg \}  
\end{eqnarray}

\begin{eqnarray}\label{ffield}
&& Z^{\mu\nu}_a  =  \frac{1+\slashed{\nu}}{2} \bigg \{P^{\mu\nu\sigma}_{3a} \gamma_5 \gamma_{\sigma} - P^{*\alpha\beta}_{2a} \sqrt{\frac{5}{3}}  \nonumber\\ && \left[g^{\mu}_{\alpha} g^{\nu}_{\beta} -\frac{g^{\nu}_{\beta}\gamma_{\alpha}\left(\gamma^\mu +\nu^\mu\right)}{5} -\frac{g^{\mu}_{\alpha}\gamma_{\beta}\left(\gamma^\nu - \nu^\nu\right)}{5}\right]\bigg \}
\end{eqnarray}


\begin{widetext}
\begin{eqnarray}
&& R^{\mu\nu\rho}_a  =  \frac{1+\slashed{\nu}}{2} \bigg \{P^{*\mu\nu\sigma}_{4a} \gamma_5 \gamma_{\sigma} - P^{\alpha\beta\tau}_{3a} \sqrt{\frac{7}{4}}  \left[g^{\mu}_{\alpha} g^{\nu}_{\beta} g^{\rho}_{\tau}   -\frac{g^{\nu}_{\beta} g^{\rho}_{\tau} \gamma_{\alpha}\left(\gamma^\mu - \nu^\mu\right)}{7}   -\frac{g^{\mu}_{\alpha} g^{\rho}_{\tau}  \gamma_{\beta}\left(\gamma^\nu - \nu^\nu\right)}{7}- \frac{g^{\mu}_{\alpha} g^{\nu}_{\beta}  \gamma_{\tau}\left(\gamma^\rho - \nu^\rho\right)}{7}\right]\bigg \}
\end{eqnarray}
\end{widetext}
where the field $H_a$ describes the $S$-wave doublet $(P, P^*)$. The $P$-wave mesons are associated with the fields $S_a$ for the $(P_0^*, P_1')$ doublet and $T_a$ for the $(P_1, P_2^*)$ doublet. The $D$-wave states are represented by the fields $X_a$ and $Y_a$, corresponding to the $(P_1^*, P_2)$ and $(P_2', P_3^*)$ doublets, respectively. Similarly, the $F$-wave meson doublets $(P_2^*, P_3)$ and $(P_3', P_4^*)$ are described by the fields $Z_a$ and $R_a$. Here, the subscript $a = u, d, s$ denotes the light flavor index.
and $\nu$ corresponds to meson four-velocity. This prescription corresponds to the ground-state mesons with quantum number $n = 1$. However, due to the heavy quark spin and flavor symmetries, heavy mesons with the same heavy quark flavor but different radial excitations ($n = 1, 2, 3, \ldots$) holds identical parity, time reversal and charge conjugation properties. These states differ in their masses only. As a result, the excited states can be incorporated into analogous effective fields: $H_a, H_a', H_a'', \ldots$, $S_a, S_a', S_a'', \ldots$ and so on, where the primes indicate higher radial excitations ($n = 2, 3, 4, \ldots$). For instance, the first radial excitation ($n = 2$) of the $S$-wave doublet is described by the field $H_a'$ representing the states $(\tilde{P}, \tilde{P}^*)$.

The light pseuoscalar mesons are incorporated by the field $\xi = e^{\frac{i \mathcal{M}}{f_\pi}}$ where the matrix $\mathcal{M}$ explicitly given as \cite{Colangelo2012}
\small{\[ \mathcal{M} = \left( \begin{array}{ccc}
\frac{1}{\sqrt{2}} \pi^0 + \frac{1}{\sqrt{6}}\eta & \pi^0 & K^+\\ \\
\pi^- & -\frac{1}{\sqrt{2}} \pi^0 + \frac{1}{\sqrt{6}}\eta & K^0\\ \\
K^- & \bar{K}^0 & -\sqrt{\frac{2}{3}}\eta \end{array} \right).\]} 
in terms of  $\pi$, $K$ and $\eta$ fields. We take $f_\pi=$ 130 MeV for calculations.  

At the leading order approximation in the heavy quark mass and light quark momentum one can construct Lagrangian in terms of these fields. The resulting Lagrangian terms denoted by $\mathcal{L}_{H}$, $\mathcal{L}_{S}$, $\mathcal{L}_{T}$, $\mathcal{L}_{X}$,
$\mathcal{L}_{Y}$, $\mathcal{L}_{Z}$ and $\mathcal{L}_{R}$ are constructed to be invariant under heavy quark spin-flavour and light quark chiral transformations as described in Ref. \cite{Falk1992}. The explicit forms of these interaction Lagrangians are given by the following expressions \cite{Colangelo2012}
\begin{equation}
\mathcal{L}_{H}= g_{H} \hspace{0.1cm} Tr\{\bar{H_a} H_b \gamma_{\mu} \gamma_{5} A^{\mu}_{ba}\}
\end{equation}

\begin{equation}
\mathcal{L}_{S}= g_{S} \hspace{0.1cm} Tr\{\bar{H_a} S_b \gamma_{\mu} \gamma_{5} A^{\mu}_{ba}\} + H.c. 
\end{equation}

\begin{equation}
\mathcal{L}_{T}= \frac{g_{T}}{\Lambda} \hspace{0.1cm} Tr\{\bar{H_a} T^\mu_b {\left(i D_\mu \slashed{A} + i \slashed{D} A_\mu \right)_{ba}} \gamma_{5} \} + H.c. 
\end{equation}

\begin{equation}
\mathcal{L}_{X}= \frac{g_{X}}{\Lambda} \hspace{0.1cm} Tr\{\bar{H_a} X^\mu_b {\left(i D_\mu \slashed{A} + i \slashed{D} A_\mu \right)_{ba}} \gamma_{5} \} + H.c. 
\end{equation}

\begin{eqnarray}\label{ylag}
\mathcal{L}_{Y}  &=& \frac{1}{\Lambda^2} \hspace{0.1cm} Tr\{\bar{H_a} Y^{\mu\nu}_b [ k^Y_1\{D_\mu, D_\nu\}  A_\lambda   \\ && + k^Y_2  \left(D_\mu D_\lambda A_\nu + \nonumber   D_\nu D_\lambda A_\mu \right) ]_{ba} \gamma^\lambda \gamma_{5} \}  + H.c.
\end{eqnarray}
\begin{eqnarray}\label{Zlag}
\mathcal{L}_{Z}  &=& \frac{1}{\Lambda^2} \hspace{0.1cm} Tr\{\bar{H_a} Z^{\mu\nu}_b [ k^Z_1\{D_\mu, D_\nu\}  A_\lambda   \\ && + k^Z_2  \left(D_\mu D_\lambda A_\nu + \nonumber   D_\nu D_\lambda A_\mu \right) ]_{ba} \gamma^\lambda \gamma_{5} \}  + H.c.
\end{eqnarray}
\begin{eqnarray}\label{Rlag}
\mathcal{L}_{R} &=&  \frac{1}{\Lambda^3} \hspace{0.1cm} Tr\{\bar{H_a} R^{\mu\nu\rho}_b [ k^R_1\{D_\mu, D_\nu, D_\rho \}  A_\lambda   +   k^R_2 \nonumber \\ && ( \{D_\mu , D_\rho\} D_\lambda A_\nu  + \{D_\mu , D_\rho\}   D_\lambda A_\nu \{D_\nu, D_\rho \} D_\lambda A_\mu  \nonumber \\ && + \{D_\mu, D_\nu \} D_\lambda A_\rho )]_{ba} \gamma^\lambda \gamma_{5} \}  + H.c. 
\end{eqnarray} \\
where the covariant derivative is defined as $D_\mu = \partial_\mu + V_\mu$; the anticommutator of two covariant derivatives is $\{D_\mu, D_\nu\} = D_\mu D_\nu + D_\nu D_\mu$ and the anticommutator between a single and a double derivative operator is $\{D_\mu, D_\nu D_\rho\} = D_\mu D_\nu D_\rho + D_\mu D_\rho D_\nu + D_\nu D_\mu D_\rho + D_\nu D_\rho D_\mu + D_\rho D_\mu D_\nu + D_\rho D_\nu D_\mu$ \cite{Colangelo2012}. The definition of vector and axial-vector currents are given as
\begin{equation}
V_\mu=\frac{1}{2}(\xi^{\dagger} \partial_\mu \xi + \xi \partial_\mu \xi^{\dagger}) 
\end{equation}
\begin{equation}
\mathcal{A}_\mu=\frac{1}{2}(\xi^{\dagger} \partial_\mu \xi - \xi \partial_\mu \xi^{\dagger})
\end{equation}
The chiral symmetry breaking scale is taken as $\Lambda =$ 1 GeV. The $g_{H}$, $g_{S}$, $g_{T}$, $g_{X}$, $g_{Y}$ = $k^Y_1 +k^Y_2$, $g_{Z}$= $k^Z_1 +k^Z_2$ and $g_{R}$= $k^R_1 +k^R_2$ are the strong coupling constants associated with the strong decays of higher excited mesons to the ground state mesons accompanied by the emission of light pseudoscalar mesons. These coupling constants can be extracted from the experimental widths. For higher excited states, the corresponding coupling constants are represented by $\tilde{g_{H}}$,  $\tilde{g_{S}}$ and so on. 

The strong decay width to $D_{(s)}^{(*)}K$ and $D_{(s)}^{(*)}\eta$ employing the chiral Lagrangians  $\mathcal{L}_{H}$, $\mathcal{L}_{S}$, $\mathcal{L}_{T}$, $\mathcal{L}_{X}$, $\mathcal{L}_{Y}$, $\mathcal{L}_{Z}$ and $\mathcal{L}_{R}$ given as~\cite{Falk1992}   
\begin{equation}
\Gamma = \frac{1}{\left(2J+1\right)} \sum \frac{p_p}{8 \pi \hspace{0.1cm} M^2_i} \hspace{0.1cm} |\mathcal{A}|^2
\end{equation}
where the magnitude of the three-momentum of the emitted pseudoscalar meson is expressed using the Källén function as
\begin{equation}
p_p = \frac{\sqrt{(M_i^2 -{(M_f +M_p)^2})\hspace{0.1cm}(M_i^2 -{(M_f -M_p)^2}})}{2 \hspace{0.1cm} M_{i}}. 
\end{equation}
Here, $M_i$ and $M_f$ are the masses of initial and final mesons, respectively; $M_p$ denotes the mass of the emitted pseudoscalar meson. The $\mathcal{A}$ represents the scattering amplitude, $J$ is the total angular momentum of the initial meson while $\Sigma$ indicates the summation over all polarization vectors of total angular momentum $j = 1, 2, 3$ or $4$. The explicit expressions for the two-body strong decay widths of heavy–light mesons into ground state pseudoscalar mesons for various decay channels are given by \cite{Colangelo1995,Colangelo2012}:

\begin{enumerate}
\item Decaying $S$ wave meson:  ${(0^-,1^-)}_\frac{1}{2}\rightarrow{(0^-,1^-)}_\frac{1}{2} + p $
 
\begin{equation}
\Gamma\left(1^- \rightarrow 0^-\right ) = \mathrm{C}_p \frac{g_{H}^2 M_f p_p^3}{6\pi f_\pi^2 M_i} 
\end{equation}

\begin{equation}
\Gamma\left(0^- \rightarrow 1^-\right ) = \mathrm{C}_p \frac{g_{H}^2 M_f p_p^3}{2\pi f_\pi^2 M_i} 
\end{equation}

\begin{equation}
\Gamma\left(1^- \rightarrow 1^-\right ) = \mathrm{C}_p \frac{g_{H}^2 M_f p_p^3}{3\pi f_\pi^2 M_i} 
\end{equation}

\item Decaying $P$ wave meson: ${(0^+,1^+)}_\frac{1}{2}\rightarrow{(0^-,1^-)}_\frac{1}{2} + p $   
 
\begin{equation}
\Gamma\left(0^+ \rightarrow 0^-\right ) = \mathrm{C}_p \frac{4 g_{S}^2 M_f}{2 \pi f_\pi^2  M_i} \left[p_p \left(M_p^2 + p_p^2\right)\right]
\end{equation} 
 
\begin{equation}
\Gamma\left(1^+ \rightarrow 1^-\right ) = \mathrm{C}_p \frac{4 g_{S}^2 M_f}{2 \pi f_\pi^2  M_i} \left[p_p \left(M_p^2 + p_p^2\right)\right]
\end{equation}
  
\item Decaying $P$ wave meson: ${(1^+,2^+)}_\frac{3}{2}\rightarrow{(0^-,1^-)}_\frac{1}{2} + p $  

\begin{equation}
\Gamma\left(2^+ \rightarrow 0^-\right ) = \mathrm{C}_p \frac{4 g_{T}^2 M_f p_p^5}{15\pi f_\pi^2 \Lambda^2 M_i} 
\end{equation}

\begin{equation}
\Gamma\left(1^+ \rightarrow 1^-\right ) = \mathrm{C}_p \frac{2 g_{T}^2 M_f p_p^5}{3 \pi f_\pi^2 \Lambda^2 M_i}
\end{equation}

\begin{equation}
\Gamma\left(2^+ \rightarrow 1^-\right ) = \mathrm{C}_p \frac{2 g_{T}^2 M_f p_p^5}{5\pi f_\pi^2 \Lambda^2 M_i} 
\end{equation}

\item Decaying $D$ wave meson: ${(1^-,2^-)}_\frac{3}{2}\rightarrow{(0^-,1^-)}_\frac{1}{2} + p $

\begin{equation}
\Gamma\left(1^- \rightarrow 0^-\right ) = \mathrm{C}_p \frac{4 g_{X}^2 M_f}{9 \pi f_\pi^2 \Lambda^2 M_i} \left[p_p^3 \left(M_p^2 + p_p^2\right)\right]
\end{equation}

\begin{equation}
\Gamma\left(1^- \rightarrow 1^-\right ) = \mathrm{C}_p \frac{2 g_{X}^2 M_f}{9 \pi f_\pi^2 \Lambda^2 M_i} \left[p_p^3 \left(M_p^2 + p_p^2\right)\right]
\end{equation}

\begin{equation}
\Gamma\left(2^- \rightarrow 1^-\right ) = \mathrm{C}_p \frac{2 g_{X}^2 M_f}{3 \pi f_\pi^2 \Lambda^2 M_i} \left[p_p^3 \left(M_p^2 + p_p^2\right)\right]
\end{equation}

\item Decaying $D$ wave meson: ${(2^-,3^-)}_\frac{5}{2}\rightarrow{(0^-,1^-)}_\frac{1}{2} + p $

\begin{equation}
\Gamma\left(2^- \rightarrow 1^-\right ) = \mathrm{C}_p \frac{4 g_{Y}^2 M_f}{15 \pi f_\pi^2 \Lambda^4 M_i} |p_p|^7
\end{equation}

\begin{equation}
\Gamma\left(3^- \rightarrow 0^-\right ) = \mathrm{C}_p \frac{4 g_{Y}^2 M_f}{35 \pi f_\pi^2 \Lambda^4 M_i} |p_p|^7
\end{equation}

\begin{equation}
\Gamma\left(3^- \rightarrow 1^-\right ) = \mathrm{C}_p \frac{16 g_{Y}^2 M_f}{105 \pi f_\pi^2 \Lambda^4 M_i} |p_p|^7 
\end{equation}

\item Decaying $F$ wave meson: ${(2^+,3^+)}_\frac{5}{2}\rightarrow{(0^-,1^-)}_\frac{1}{2} + p $

\begin{equation}
\Gamma\left(2^+ \rightarrow 0^-\right ) = \mathrm{C}_p \frac{4 g_{Z}^2 M_f}{25 \pi f_\pi^2 \Lambda^4 M_i} \left[p_p^5 \left(M_p^2 + p_p^2\right)\right]
\end{equation}

\begin{equation}
\Gamma\left(2^+ \rightarrow 1^-\right ) = \mathrm{C}_p \frac{8 g_{Z}^2 M_f}{75 \pi f_\pi^2 \Lambda^4 M_i} \left[p_p^5 \left(M_p^2 + p_p^2\right)\right]
\end{equation}

\begin{equation}
\Gamma\left(3^+ \rightarrow 1^-\right ) = \mathrm{C}_p \frac{4 g_{Z}^2 M_f}{15 \pi f_\pi^2 \Lambda^4 M_i} \left[p_p^5 \left(M_p^2 + p_p^2\right)\right]
\end{equation}

\item Decaying $F$ wave meson: ${(3^+,4^+)}_\frac{7}{2}\rightarrow{(0^-,1^-)}_\frac{1}{2} + p $

\begin{equation}
\Gamma\left(4^+ \rightarrow 0^-\right ) = \mathrm{C}_p \frac{16 g_{R}^2 M_f p_p^9}{35 \pi f_\pi^2 \Lambda^6 M_i}
\end{equation}

\begin{equation}
\Gamma\left(3^+ \rightarrow 1^-\right ) = \mathrm{C}_p \frac{36 g_{R}^2 M_f p_p^9}{35 \pi f_\pi^2 \Lambda^6 M_i}
\end{equation}

\begin{equation}
\Gamma\left(4^+ \rightarrow 1^-\right ) = \mathrm{C}_p \frac{4 g_{R}^2 M_f p_p^9}{7 \pi f_\pi^2 \Lambda^6 M_i}
\end{equation}

\end{enumerate}
The coefficient $\mathrm{C}_p$ appearing in the decay width expressions takes different values depending on the emitted pseudoscalar meson: $\mathrm{C}_{\pi^\pm} = \mathrm{C}_{K^\pm} = 1$, $\mathrm{C}_{\pi^0} = \frac{1}{2}$, $\mathrm{C}_\eta = \frac{2}{3}$ \cite{Colangelo2012}. The strong coupling constants and their notations depend on the radial quantum numbers involved in the transition. For transitions with $n_i = n_f = 1$, the couplings are denoted as $g_{H}, g_{S}, g_{T}, g_{X}, g_{Y}, g_{Z}, g_{R}$ whereas for transitions from $n_i = 2$ to $n_f = 1$, they are represented as $\Tilde{g}^2_{H,S,T,X,Y,Z,R}$. Higher order loop corrections are neglected to avoid the introduction of additional coupling constants. In the present analysis, we employ the approximation $\mathcal{A}_\mu = i \frac{\partial_\mu \mathcal{M}}{f_\pi}$. We note that, when the momenta of the emitted light pseudoscalar mesons are not significantly small, additional terms may be included to account for higher order effects, which however introduce new undetermined coupling constants. Furthermore, spin and flavor violating corrections arising at order $\mathcal{O}(1/m_Q)$ in the heavy quark expansion can become non negligible. In such cases, the introduction of extra coupling constants may affect the predictions and may not fully cancel in ratios of decay widths. Nevertheless, these subleading contributions are expected to remain significantly smaller than the leading order terms and are thus neglected in this work. The leading order unknown coupling constants that appear in decay rates can either be theoretically estimated or extracted from available experimental data on decay widths. Successful predictions of these couplings have been reported in the literature using approaches such as QCD sum rules \cite{Colangelo1995} and lattice QCD \cite{Becirevic2014}. The numerical values of the meson masses used as input parameters for present decay width calculations are $M_{D^{*+}} = 2010.26 $ MeV, 
$M_{D^{*0}} = 2006.85 $ MeV, 
$M_{D^{+}} = 1869.65 $ MeV, 
$M_{D^{0}} = 1864.83 $ MeV, 
$M_{\pi^{0}} = 134.97 $ MeV, 
$M_{\pi^{-}} = 139.57 $ MeV,  $M_{K^{-}} = 493.67 $ MeV, 
$M_{\eta} = 547.86 $ MeV \cite{PDG2020}.

\begin{table*}
\begin{center}
\tabcolsep 15pt
\caption{{The strong decay widths (in MeV) for charm-strange resonance  with possible spin-parity assignments. The ratio is calculated from $\frac{\Gamma}{\Gamma(D_{sJ} \rightarrow D^{*}K)}$. Fraction (in $\%$) represents the percentage of the partial decay width with respect to total decay width. }}\label{all}
\begin{tabular}{lclrccc}
\hline\hline\noalign{\smallskip}
Resonance & State $(_{s_l} J^P)$   & Decay Mode & Width & Ratio & Fraction & Exp (in MeV) \\
\noalign{\smallskip}\hline\noalign{\smallskip}
$D_{s2}^*(2573)$ & $1^3P_2$ $(_\frac{3}{2} 2^+)$   & D$^{*+} K^{0}$ &  2.42 $g_T^2$\\
 &  &   D$^{*0} K^{+}$ &  3.33 $g_T^2$ \\
 &  &   D$^{+} K^{0}$ &  44.06 $g_T^2$ \\
 &  &   D$^{0} K^{+}$ &  48.92 $g_T^2$\\
 &  &   D$^{*} K$ & 2.87  $g_T^2$ & 1 & 4.97 &\\
 &  &   D$ K$ & 46.49  $g_T^2$ & 16.20 & 80.37 &\\ 
 &  &   D$^{*}_s \eta$ &  7.67 $g_T^2$ & 2.67 & 13.26 & \\
 &  &   D$_s \eta$ &  0.81 $g_T^2$ & 0.28 & 1.40 & \\
 &  &    Total & 57.84 $g_T^2$ & & & $16.9 \pm 0.8$ \\
$D_{sJ}(2700)$ & $2^3S_1$ $(_\frac{1}{2} 1^-)$   & D$^{*+} K^{0}$ &  422.91 $\tilde{g}_H^2$\\
&  &   D$^{*0} K^{+}$ &  442.04$\tilde{g}_H^2$ \\
&  &   D$^{+} K^{0}$ &  439.50 $\tilde{g}_H^2$ \\
&  &   D$^{0} K^{+}$ &  453.40 $\tilde{g}_H^2$\\
 &  &   D$^{*} K$ &  432.47 $\tilde{g}_H^2$ & 1 & 40.29 & \\
 &  &   D$ K$ & 446.45 $\tilde{g}_H^2$ & 1.03 & 41.60 &\\
 &  &   D$^{*}_s \eta$ &  53.28 $\tilde{g}_H^2$ & 0.12 & 4.96 & \\
 &  &   D$_s \eta$ &  141.08 $\tilde{g}_H^2$ & 0.32 & 13.14 & \\
 &  &    Total & 1073.28 $\tilde{g}_H^2$ & & & $115.8 \pm 7.3$ \\
$D_{sJ}(2700)$ & $1^3D_1$ $(_\frac{3}{2} 1^-)$   & D$^{*+} K^{0}$ &  260.93 ${g}_X^2$\\
 &  &   D$^{*0} K^{+}$ &  270.63 ${g}_X^2$ \\
 &  &   D$^{+} K^{0}$ &  1155.79 ${g}_X^2$ \\
 &  &   D$^{0} K^{+}$ &  1191.27 ${g}_X^2$\\
 &  &   D$^{*} K$ & 265.33  ${g}_X^2$ & 1 & 11.39 &  \\
 &  &   $ DK$ & 1173.53 $g_X^2$ & 4.42 & 50.40 &\\
 &  &   D$^{*}_s \eta$ &  13.87 ${g}_X^2$  & 0.05 & 0.60 &\\
 &  &   D$_s \eta$ &  875.81 ${g}_X^2$ & 3.30 & 37.61 &\\
 &  &    Total & 2328.55 ${g}_X^2$  & & & $115.8 \pm 7.3$ \\
$D_{s0}(2590)$ & $2^1S_0$ $(_\frac{1}{2} 0^-)$   & D$^{*+} K^{0}$ &  184.67 $\tilde{g}_H^2$\\
 &  &   D$^{*0} K^{+}$ &  204.11 $\tilde{g}_H^2$ \\
 &  &   D$^{*} K$ & 194.39  $\tilde{g}_H^2$ & 1 & $\sim$ 100 & \\
 &  &    Total  & 194.39 $\tilde{g}_H^2$ & & & $89 \pm 16$ \\
 \noalign{\smallskip}\hline\hline
    \end{tabular}%
 \end{center}
\end{table*}



\begin{table*}
\begin{center}
\tabcolsep 25pt
\caption{{The strong decay widths (in MeV) for resonance $D_{sJ}(2860)$ with possible spin-parity assignments. The last raw shows ratio $\frac{D^*K}{DK}$ for possible assignments independent of coupling constant.}}\label{2860s3ds}
\begin{tabular}{lccccc}
\hline\hline\noalign{\smallskip}
 & $2^3P_2({_\frac{3}{2}} 2^+)$ & $1^3D_1({_\frac{3}{2}} 1^-)$ & $1^3D_3({_\frac{5}{2}} 3^-)$   & $1^3F_2({_\frac{5}{2}} 2^+)$   \\
Decay Mode &  (in $\tilde{g}_T^2$ MeV) & (in $g_X^2$ MeV)  & (in in $g_Y^2$ MeV) &   (in $g_Z^2$ MeV)\\
\hline\noalign{\smallskip}
D$^{*+} K^{0}$ & 1320.31 & 260.63 & 43.29 & 994.34  \\
D$^{*0} K^{+}$ & 1356.55 & 270.04 & 46.25 & 1015.68   \\
D$^{+} K^{0}$ & 1545.20 & 1155.79 & 114.77 & 2595.11  \\
D$^{0} K^{+}$ & 1584.76 & 1191.27 & 121.30 & 2652.63   \\

D$^{*} K$ & 1338.43 & 265.33 & 44.77 & 1005.01  \\
D$ K$ & 1564.98 & 1173.53 & 118.03  & 2623.87    \\
D$^{*}_s \eta$ & 422.66 & 13.87 & 3.99 & 3298.2  \\
D$_s \eta$ & 609.72  & 875.81 & 22.91 & 9852.47 \\
Total & 3935.79 & 2328.55 &  189.72 & 16779.55 \\
$\frac{D^*K}{DK}$ & 0.85 & 0.22 & 0.37 & 0.38 \\
\noalign{\smallskip}\hline\hline  
\end{tabular}
\end{center}
\end{table*}

\begin{table}
\begin{center}
\tabcolsep 2pt
\caption{{The strong decay widths (in MeV) for $1^3D_2$ and $1^1D_2$ states. Fraction (in $\%$) represents the percentage of the partial decay width with respect to total decay width.}}\label{spinpartners}
\begin{tabular}{lcccc}
\hline\hline\noalign{\smallskip}
 & $1^3D_2({_\frac{5}{2}} 2^-)$ & & $1^1D_2({_\frac{3}{2}} 2^-)$  &   \\
Decay Mode &  (in $g_Y^2$ MeV) & Fraction & (in ${g}_X^2$ MeV)  & Fraction\\
\hline\noalign{\smallskip}
D$^{*+} K^{0}$ & 57.73 &  & 956.03 &  \\
D$^{*0} K^{+}$ & 61.99 & & 987.48 &   \\
D$^{*} K$ & 59.86 & 93.24  & 971.75 & 80.81  \\
D$^{*}_s \eta$ & 4.34 & 6.76 & 230.87 & 19.19\\
 Total & 16.05 & & 81.30 & \\
\noalign{\smallskip}\hline\hline  
\end{tabular}%
\end{center}
\end{table}%

\begin{table*}
\begin{center}
\tabcolsep 13pt
\caption{{The strong decay widths (in MeV) for resonance $D_{SJ}(3040)$ with possible spin-parity assignments.}}\label{3040ds}
\begin{tabular}{lccccc}
\hline\hline\noalign{\smallskip}
 & $2^1S_0({_\frac{1}{2}} 0^-)$ &  $2^1P_1({_\frac{1}{2}} 1^+)$   & $2^3P_1({_\frac{3}{2}} 1^+)$  & $1^1D_2{(_\frac{3}{2}} 2^-)$ & $1^3D_2({_\frac{5}{2}} 2^-)$ \\
Decay Mode &  (in $\tilde{{g}}_H^2$ MeV) &   (in $\tilde{g}_S^2$ MeV) &   (in $\tilde{g}_T^2$ MeV) &   (in $g_X^2$ MeV)&   (in $g_Y^2$ MeV)\\
\hline\noalign{\smallskip}
D$^{*+} K^{0}$ & 184.66  & 14448.9 & 1858.00 & 956.03 & 57.33\\
D$^{*0} K^{+}$ & 204.12  & 14567.8 & 1912.72 & 987.48 & 61.99 \\
D$^{*} K$ & 194.39  & 14508.35 & 1885.36 & 971.75 & 59.86 \\
D$^{*}_s \eta$ & \ldots  & 7397.7 & 560.76 & 230.87 & 4.34\\

 Total & 194.39 & 21906.05 & 2446.12 & 1202.63 & 64.21 \\
\noalign{\smallskip}\hline\hline  
\end{tabular}%
\end{center}
\end{table*}


\begin{table}
\begin{center}
\tabcolsep 18pt
\small
\caption{\label{23p0}%
The strong decay widths (in MeV) for  \small{$2^3P_0$} state. Fraction (in $\%$) represents the percentage of the partial decay width with respect to total decay width.}
\begin{tabular}{lcc}
\hline\hline
&  $2^3P_0 (_\frac{1}{2} 0^+)$ & \\  
Decay Mode & (in $\tilde g_S^2$ MeV)   &  Fraction   \\
\hline
$D^{+}K^0$ & 18317.60 &   \\
$D^0K^+$ & 18469.70  &      \\
$DK$ & 18393.65 &  87.02  \\
$D_s\eta$ & 2742.65  &   12.97  \\
Total & 211.36  &    \\
\hline\hline 
\end{tabular}    
\end{center}
\end{table}

\section{Results and Discussion}\label{RD}
Using the present relativistic formalism, the calculated mass spectra of charm–strange mesons ($c\bar{s}$) for ground, radial and orbitally excited states are presented in Tables \ref{swds}–\ref{dfwds}. These predictions are obtained using the model parameters summarized in Table \ref{input}. The mesons $D_s^*$ (2112.2 $\pm$ 0.4 MeV) and $D_s$ (1968.35 $\pm$ 0.07 MeV) are well established as the ground states of the charm–strange sector \cite{PDG2020}. In present study, the masses of the $1^3S_1$ and $1^1S_0$ states are predicted to be 2113.14 MeV and 1968.95 MeV, respectively. These values show excellent agreement with the corresponding experimental results reported by the PDG \cite{PDG2020}. {{Incorporating the charm quark mass uncertainty, $m_c = 1.27 \pm 0.02$ GeV (~1.5$\%$) as reported by the PDG \cite{PDG2020}, results in variations of approximately 0.9$\%$ and 1.0$\%$ in the predicted masses of the $1^3S_1$ and $1^1S_0$ states of the charm-strange meson, respectively. The strange quark mass, being relatively small contributes negligibly to the overall uncertainty. Variations of the potential model parameters $\lambda$, $V_0$ and $\sigma$ by 5–10$\%$ result in shifts of up to 2$\%$ in the predicted mass of the $1S$ state, with $\lambda$ exhibiting the highest sensitivity. The quantitative estimates of these uncertainties and the methodology used to obtain them can be found in Appendix B of the previous work \cite{pandya2021}.}}

The states $D_{s1}(2536)$ and $D_{s2}(2573)$ are well-established members of the $1P_{3/2}$ multiplet, corresponding to the $(1^3P_1, 1^3P_2)$ states of charm–strange mesons. Despite of having lower masses, the $D_{s0}^*(2317)$ and $D_{s1}(2460)$ are listed as the $1P_{1/2}$ multiplet $(1^3P_0, 1^1P_1)$ in the PDG listings \cite{PDG2020}. The observed mass gap between the $1P_{3/2}$ and $1P_{1/2}$ doublets cannot be fully explained using spin-dependent interactions alone suggesting a possible exotic nature for the $1P_{1/2}$ states. In the present study, the predicted masses for the $D_{s1}(2536)$ and $D_{s2}(2573)$ states are found to be 2517.25 MeV and 2561.41 MeV, respectively, which show good agreement with the PDG reported experimental averages of $2535.11 \pm 0.06$ MeV and $2569.1 \pm 0.8$ MeV. The consistency of our predictions for both the $1S_{1/2}$ and $1P_{3/2}$ multiplets with experimental data provides a strong basis for assigning quantum numbers $J^P$ to newly observed excited states reported by BaBar, Belle and LHCb. However, to provide a meaningful description of the charm–strange meson states, predictions of the $D^{*}K/DK$ decay ratios are also essential for testing the effectiveness of the present formalism in the context of strong decays. While experimental branching ratio data for most $1P$ charm–strange states remain inadequate, the ratio for $D^*_{s2}(2573)$ has been measured and is listed by PDG as \cite{PDG2020}
\begin{equation}\label{2573ratio}
 {\frac{\mathcal{B}(D^*_{s2}(2573) \rightarrow D^{*} K) }{\mathcal{B}(D^*_{s2}(2573) \rightarrow D K)}} < 0.33   
\end{equation}
Considering $D^*_{s2}(2573)$ as a well established $1^3P_2$ state, its OZI-allowed strong decays to ground state mesons with the emission of a light pseudoscalar meson are governed by the coupling constant $g_T$. The partial decay widths of this state, expressed in terms of $g_T$, are presented in Table \ref{all}. The resulting decay ratio is $
\frac{\Gamma(D^*_{s2}(2573) \rightarrow D^* K)}{\Gamma(D^*_{s2}(2573) \rightarrow D K)} = 0.061 $ which lies well within the experimental upper limit and also compatible with the value $0.086$ quoted using HQET in Ref. \cite{Colangelo2012}.
This consistency therefore further validates the reliability of our formalism in describing strong decay processes. Here, the notation $D^{(*)}K$ represents the isospin averaged decay channels $D^{(*)0}K^+ + D^{(*)+}K^0$. Additionally, we predict the ratio $\frac{\Gamma(D^*_{s2}(2573) \rightarrow D^{*0} K^+)}{\Gamma(D^*_{s2}(2573) \rightarrow D^0 K^+)} = 0.067 $ which can serve as a useful input for future experimental investigations. From the present study, we found $g_T = 0.52 \pm 0.082$ which is consistent with values obtained in earlier studies, such as $g_T = 0.43 \pm 0.05$ \cite{Wang2014} and $0.43 \pm 0.01$ \cite{wang20131}. {{The procedure for estimating the uncertainties associated with the coupling constants can be found in Appendix C of our previous work \cite{pandya2021}}}.

\subsubsection{$D_{SJ}(2860)^{+}$}
In 2006, the $BABAR$ collaboration observed a new state around the mass region of 2860 MeV, designated as $D_{SJ}(2860)^{+}$, in the inclusive $DK$ channel \cite{BABAR2006DS}. Based on their spin analysis and the absence of decay to final states such as $D^*K$ or $D^*_s \eta$, the $BABAR$ collaboration concluded that $D*{SJ}(2860)^{+}$ could be explained with natural parity quantum numbers as $0^+, 1^-, 2^+, 3^-, ldots$, etc. Thus, the possible spin-parity assignments for this state include  $2^3S_1$, $2^3P_0$, $2^3P_2$, $1^3D_3$, $1^3D_1$ and $1^3F_2$. The properties and mass range of the state exclude its identification as a radially excited $2S$ state. Three years later, the $BABAR$ collaboration reported the decay of $D_{SJ}^*(2860)^{+}$ to the $D^*K$ final state along with a measurement of branching fractions relative to the $DK$ final state \cite{BABAR2009ds}. This suggests that$D_{SJ}(2860)^+$ cannot be assigned $J^P=0^+$, as the decay of a $^3P_0$ state to $D^*K$ is forbidden. Consequently, the $2^3P_0$ assignment for $D_{SJ}^*(2860)^{+}$ is excluded. From the predicted masses obtained in the present study, $1^3D_1$ (2819.57 MeV), $1^3D_3$ (2853.86 MeV), $2^3P_2$ (3067.23 MeV) and $1^3F_2$ (3277.45 MeV) it is evident that the mass values of $2^3P_2$ and $1^3F_2$ lie significantly above the average mass of 2860.5 $\pm$ 2.6 $\pm$ 6.5 MeV reported by the PDG for $D_{sJ}(2860)$. Based on mass considerations, $D_{sJ}(2860)$ is therefore most likely to correspond to either the $1^3D_1$ or $1^3D_3$ state.

Furthermore, using our computed mass values, the partial decay widths expressed in terms of effective coupling constants for four possible assignments of  $D_{sJ}(2860)^{+}$ are presented in Table \ref{2860s3ds}. In all considered cases, the $DK$ mode appears to be the most dominant. However, the $2^3P_2$ assignment can be ruled out as the $DK$ mode is less dominant compared to other decay channels. Similarly, the possibility of $D_{sJ}(2860)^{+}$ being a $1^3F_2$ state is also excluded as it results in an excessively large total decay width inconsistent with experimental observations. The two remaining possibilities $1^3D_1$ and $1^3D_3$, exhibit similar decay patterns but differ significantly in their predicted widths. We find that the $1^3D_3$ state yields a relatively narrow decay width of 189.72 $g_Y^2$ whereas the $1^3D_1$ state produces a much broader width of 2328.55 $g_X^2$. The experimentally measured total width of $D_{sJ}(2860)^{+}$ as reported by both $BABAR$ and $LHCb$ lies in the range of 48 to 69 MeV with only a few MeV of uncertainty. Therefore, based on its mass and decay characteristics, we assign the $D_{sJ}(2860)$ state observed by the $BABAR$ and $LHCb$ collaborations as the $1^3D_3$ charm-strange meson. Furthermore, identifying $D_{sJ}(2860)$ as the $1^3D_3$ state and comparing the computed width $189.72 \hspace{0.1cm} g_Y^2$ with the experimentally measured width of $48 \pm 3$ MeV, we extract the coupling constant $g_Y = 0.50 \pm 0.015$. This value is in close agreement with previously reported estimates of $0.49 \pm 0.039$ \cite{pandya2021}, $0.53 \pm 0.13$ \cite{Wang2014} and $0.42 \pm 0.02$ \cite{wang20131}.

The ratio 
\begin{equation} 
\frac{\mathcal{B}(D_{sJ}^*(2860) \rightarrow D^* K)}{\mathcal{B}(D_{sJ}^*(2860) \rightarrow D K)} = 1.10 \pm 0.15 \pm 0.19
\end{equation}
has been measured by the $BABAR$ collaboration \cite{BABAR2009ds}. In the present study, this ratio is obtained as 0.38 which closely aligns with the prediction of 0.39 by Colangelo \cite{Colangelo2006} and 0.38 obtained by Wang \cite{wang2015} within the framework of HQET. However, this theoretical estimates are considerably lower than the experimental value reported by $BABAR$ \cite{BABAR2009ds}. The $^3P_0$ model predicts this ratio to lie in the range 0.55 to 0.80 \cite{Song2015,li2010}, while other theoretical quark models suggest values between 0.43 and 0.71 \cite{godfrey20141,godfrey20142,Segovia2015}. As we note from the charm sector, the $D_3^*(2760)$ state is regarded as the charm counterpart of the $D_{s3}^*(2860)$. Additionally, the existence of the $D_2(2740)$ state with $J^P = 2^-$, located close in mass to the $J^P = 3^-$ state, suggests a similar pattern may occur in the charm-strange sector. Given the comparable mass gap between the charm and charm-strange spectra, it is reasonable to expect a nearby $J^P = 2^-$ state around the mass of $D_{s3}(2860)$. This proximity could potentially influence the observed decay channels and modify the measured branching ratio reported by experiment.

In 2014, the $LHCb$ collaboration conducted a Dalitz plot analysis of the decay $B_s^0 \rightarrow \bar{D}^0 K^- \pi^+$, revealing two distinct resonant structures near 2.86 GeV in the $\bar{D}^0 K^-$ invariant mass distribution identified as $D_{s1}^*(2860)$ and $D_{s3}^*(2860)$ with a statistical significance exceeding 10$\sigma$ \cite{lhcb2014dsprl}. The observation of these two nearby states with differing spin-parity assignments strongly supports their interpretation as members of the $1D$ charm-strange meson multiplet, with $D_{s3}(2860)$ corresponding to the $1^3D_3$ state and $D_{s1}(2860)$ to the $1^3D_1$ state. Furthermore, our analysis indicates that the resonance exhibiting a comparatively broader decay width (2328.55 $g_X^2$) is consistent with the expected decay pattern of a $1^3D_1$ state, thereby strengthening its identification as the $1^3D_1$ member of the spectrum. By comparing the experimentally measured width of $159 \pm 23$ MeV for $D_{s1}(2860)$ with the our prediction of $2328.55\, g_X^2$, we obtained the effective coupling constant as $g_X = 0.26 \pm 0.018$, which is in good agreement with the value $0.19 \pm 0.04$ reported by Wang \cite{Wang2014}.

Following the identification of $D_{s3}(2860)$ and $D_{s1}(2860)$ as the $1^3D_3$ and $1^3D_1$ states, it becomes essential to get information about their corresponding spin-partner states, $1^3D_2$ and $1^1D_2$, which are yet to be observed experimentally. Our predicted masses for these missing states are 2830.71 MeV and 2848.63 MeV, respectively. The OZI allowed decay modes for both states are presented in Table \ref{spinpartners}. Using the extracted values $g_Y = 0.50 \pm 0.015$ and $g_X = 0.26 \pm 0.018$, we estimate their total decay widths to be 16.05 MeV for the $1^3D_2$ and 81.30 MeV for the $1^1D_2$. These results suggest that both states are expected to exhibit narrower decay widths than their spin counterparts, $1^3D_3$ and $1^3D_1$,which may influence the feasibility of their detection in experiments.

\subsubsection{$D_{sJ}(3040)$}
The first evidence for the $D_{sJ}(3040)$ state was reported by the $BABAR$ collaboration through its observation in the $D^*K$ decay channel angular analysis \cite{BABAR2009ds}. This indicates that the state possesses an unnatural parity allowing for possible quantum number assignments such as $J^P = 0^-, 1^+, 2^-, \dots$. Theoretically allowed states corresponding to these quantum numbers include $2^1S_0$, $2^3P_1$, $2^1P_1$, $1^3D_2$, $1^1D_2$. The predicted mass for the $2^1S_0$ state is 2609.73 MeV, which is approximately 400 MeV lower than the experimental mass of $3044 \pm  8^{+30}_{-5}$ MeV reported by reported by $BABAR$ for $D_{sJ}(3040)$ state. Moreover, in present study the states $D_{s1}(2860)$ and $D_{s3}(2860)$ have been identified as the $1^3D_1$ and $1^3D_3$ states, respectively. A mass gap of approximately 200 MeV is too large to be attributed to closely spaced states within the same $1D$ multiplet, thus excluding the possibility of $D_{sJ}(3040)$ being $1^3D_2$ or $1^1D_2$ state. Similarly, the $2D$ states are predicted to be around 3400 MeV in the present study, which is considerably higher than the observed mass of the $D_{sJ}(3040)$ thus ruling out these assignments. This narrows the plausible assignments to the $2^3P_1$ and $2^1P_1$ states.  We predict the mass of $2^3P_1$ as 3035.06 MeV whereas mass of $2^1P_1$ as 3027.09 MeV. Both of the assignments leads to $J^P = 1^+$ for $D_{sJ}(3040)$ state.

The calculated partial widths in terms of the relevant coupling constant for possible spin-parity assignments of $D_{sJ}(3040)$ are presented in Table \ref{3040ds}. The $2^1S_0$, $1^3D_2$ and $1^1D_2$ states can also be excluded since their predicted total decay widths found to be significantly narrow than the broad width $239 \pm 35^{+40}_{-42}$ MeV measured for the $D_{sJ}(3040)$. Among the remaining candidates, $2^1P_1$ and $2^3P_1$, the $D_s^*\eta$ decay mode is less dominant for the $2^3P_1$ state. Moreover, the predicted total decay width of the $2^3P_1$ state is significantly narrower than the experimentally observed width of the $D_{sJ}(3040)$, apparently ruling out this assignment. The effective coupling constants are considered to be approximately universal across heavy meson sectors within the framework of heavy quark symmetry and effective Lagrangian approaches. This universality allows for couplings determined in one flavor sector to be used reliably as reference inputs for making predictions in another sector. Accordingly, for the $2^1P_1$ assignment of the $D_{sJ}(3040)$, employing the value $\tilde{g}_s = 0.11 \pm 0.015$ as extracted in our earlier study of charm mesons \cite{pandya2021}, the total decay width of the $D_{sJ}(3040)$ is estimated to be 265.06 MeV. This is in reasonable agreement with the experimentally measured width of $239 \pm 35^{+40}_{-42}$ MeV especially when the sizeable uncertainties are taken into account. We note that the prediction for the coupling $\tilde{g}_T$ is not presently available in the literature, however it is theoretically anticipated that $\tilde{g}_T < g_T$ due to flavor symmetry breaking and associated dynamical phase space suppression effects. As a result, the total width for the $D_{sJ}(3040)$ as $2^3P_1$ assignment will be less than measured broad width. These considerations provide additional support to the consistency of our theoretical predictions and strengthens the interpretation of the $D_{sJ}(3040)$ as a $2^1P_1$ state. Pierror and Eitchen has predicted the large width of $\sim$ 256 MeV for $2^1P_1$ state and narrow width of $\sim$ 58 MeV for $2^3P_1$ state based on chiral quark model \cite{Pierro2001}. For $D_{sJ}(3040)$ state, similar interpretations are also given by $^3P_0$ model \cite{Sun2009} and flux tube model \cite{chen2009}. {{We note the fact that heavy-light mesons are not eigenstates of the charge conjugation operator, resulting in significant mixing between states with identical $J^P$ quantum numbers via spin-orbit interactions which is typically described by a mixing angle near $35.3^\circ$. Consequently, the admixture of $2^3P_1$ and $2^1P_1$ states, as discussed in Refs. \cite{Xiao2014, Song2015}, cannot be neglected for the state  $D_{sJ}(3040)$. However, within the HQET the coupling constants for the pure $n^3P_1$ and $n^1P_1$ configurations differ substantially, so it is justified not to include mixing effects in the present work to avoid the introduction of additional complex parameters.}} 

From present analysis, the effective coupling constant $\tilde{g}_S$ determined to be  0.10 $\pm$ 0.010 deduced by equating the theoretical total width of 21905.75 $\tilde{g_S}^2$ for $D_{sJ}(3040)$ with the experimentally measured width of $239 \pm 35$ MeV. The state $2^1P_1$ has corresponding spin partner within the same doublet as the $2^3P_0$ state. Our predicted mass of the $2^3P_0$ state is 3017.41 MeV and its OZI allowed decay widths and branching fractions are listed in Table \ref{23p0}. Among the accessible decay modes, the $DK$ channel emerges as the most promising for the future experimental observation of the $D_s(2^3P_0)$ state.

\subsubsection{$D_{sJ}(2700)$}
The $D_{sJ}(2700)$ resonance was first observed by the $BABAR$ collaboration in $D^0K^+$ and $D^+K_s^0$ invariant mass distribution with mass and width reported as 2688 $\pm$ 4 $\pm$ 3 MeV and 112 $\pm$ 7 $\pm$ 36 MeV, respectively \cite{BABAR2006DS}. This state was then independently confirmed by $Belle$ collaboration in the Dalitz plot of $B^+ \rightarrow \bar{D}^0 D^0 K^+$ process with $J^P = 1^-$ \cite{belle2008ds}. The $D^*K$ mode observed by $BABAR$ for this state along with the branching ratio relative to $DK$ mode as \cite{BABAR2009ds}
\begin{equation}\label{2700ratio}
\frac{\mathcal{B}(D_{s1}^*(2710) \rightarrow D^* K)}{\mathcal{B}(D_{s1}^*(2710) \rightarrow D K)} = 0.91 \pm 0.13 \pm 0.12
\end{equation}
The $LHCb$ also confirm this state in $DK$ mass distribution consistent with total angular momentum 1 and negative parity \cite{lhcb2012ds}. Within this quantum number assignment $J^P = 1^-$, the possible candidates in the mass range of 2700–2800 MeV are the $2^3S_1$ and $1^3D_1$ states. Our predicted mass of $2^3S_1$ as 2731.51 MeV and that of $1^3D_1$ is 2819.57 MeV, then supports the assignment of $D_{sJ}(2700)$ as the radial excited $2^3S_1$ state. The OZI allowed partial decay widths corresponding to both probable assignments are summarized in Table \ref{all}. In both the $2^3S_1$ and $1^3D_1$ assignments, the $DK$ channel is found to be dominant. However, for the $2^3S_1$ state, the decay contributions from both $DK$ and $D^*K$ modes are nearly similar which aligns well with experimental observations. We notice that the $1^3D_1$ state exhibits a significantly enhanced branching ratio for the $D_s \eta$ mode along with an anomalously large branching fraction for $DK$ which is not supported by the available experimental data for the resonance $D_{sJ}(2700)$. Also, the mass range around 2800-2860 was not observed by $Belle$ collaboration in $B^+ \rightarrow \bar{D}^0 D^0 K^+$ analysis with 414 $fb^{-1}$ integrated luminosity and $449 \times 10^{6}$ $B\bar{B}$ pairs at $\sqrt{s} = 10.58$ GeV \cite{belle2008ds}.  The suppression of production around the mass range 2800-2860 in $B$ decay processes can be justified by high spin and higher excitation in charm-strange sector \cite{belle2008ds}. Therefore, by taking into account both the mass and the partial strong decay widths, we confirm the $D_{sJ}(2700)$ as the $2^3S_1$ state. Moreover, with this assignment, the ratio given in Eq. (\ref{2700ratio}) obtained as 0.96, which is in excellent agreement with the experimental measurement of $0.91 \pm 0.13 \pm 0.12$ \cite{BABAR2009ds,PDG2020}.

Furthermore, the total decay width of $D_{sJ}(2700)$ is obtained as 1073.30 $\tilde{g_H}^2$. The extracted value of the coupling constant $\tilde{g_H}$ as $0.32 \pm 0.010$ is in excellent agreement with $0.31 \pm 0.017$ from our previous study \cite{pandya2021} and $0.28 \pm 0.010$ reported in Ref. \cite{wang20131}. However, it is worth noting that these values are almost double the value of $0.14 \pm 0.030$ obtained in Ref. \cite{Wang2014}. This could be explain by the fact that the radial excited state can further decay to $1P$ state with the emission of the light pseudoscalar mesons $(\pi, \eta, K)$. Taking account those channels necessitates the inclusion of the unknown coupling constants to the theory. In order to prevent complexities, we do not consider them here. The $D_{sJ}(2700)$ as $2^3S_1$ assignment has been supported by the study based on HQET \cite{Campanella2018} and by the relativistic quark model based on quasi potential approach \cite{Ebert2010}. Other plausible interpretation considers $D_{sJ}(2700)$ as a mixed  $2^3S_1$$-$$1^3D_1$ state. This possibility has been explored in various studies including constituent quark model \cite{zhong2010}, the model assuming canonical $c\bar{s}$ states  \cite{close2007} and $^3P_0$ model  \cite{li2010,Song2015}. Additionally, interpretations of $D_{sJ}(2700)$ as a charmed tetraquark state \cite{wang2008} or as a hadronic molecular configuration have also been proposed \cite{vinodkumar2008}. 

The spin partner of the $2^3S_1$ state is the pseudoscalar $2^1S_0$ state which has not yet been firmly established in the charm-strange meson spectrum. Recently, $LHCb$ has observed a new state labelled as $D_{s0}(2590)$ with mass of 2591 $\pm$ 6 $\pm$ 7 MeV and width of 89 $\pm$ 16 $\pm$ 12 MeV in $B^0 \rightarrow D^-D^+K^+{\pi^-}$ channel decaying to $D^+K^+{\pi^-}$ final state \cite{lhcb2021ds}. Below $K^*(892)^0$ threshold, the $K^+\pi^-$ system belongs to $S$ wave only. This allows only unnatural parity states $(J^P= 0^-, 1^+, 2^-, \ldots)$ decay to $D^+K^+\pi^-$. In the present study, the predicted mass of the $2^1S_0$ state is 2629.11 MeV. Moreover, the mass difference between the state $D_{SJ}(2700)$ and newly reported $D_{s0}(2590)$ is approximately 112 MeV which is comparable to the mass difference of 122 MeV between the states $D^*(2600)$ and $D(2550)$ \cite{pandya2021}. Given these characteristices, $D_{s0}(2590)$ state emerges as a strong candidate for first radial excitation of the pseudoscalar charm-strange meson.

The partial widths of OZI allowed strong decays for the $D_{s0}(2590)$ state as $2^1S_0$ candidate are shown in Table \ref{all}. The total decay width is found to be 194.39 $\tilde{g_H}^2$ MeV which corresponds to 19.90 MeV for $\tilde{g_H} = 0.32 \pm 0.010$. This predicted width is considerably lower than the experimentally measured width of $89 \pm 16 \pm 12$ MeV even for the lower limit. However, the predicted width aligns reasonably well with the predictions from the chiral quark model \cite{Ni2022} and $^3P_0$ model \cite{Xie2021}. Due to this discrepancy, a conclusive identification of the $D_{s0}(2590)$ as the pseudoscalar radial excitation remains uncertain. More experimental efforts are then required to improve our understanding of this state. Overall, the combined analysis of mass predictions and strong decay behavior presented in this work offers valuable theoretical input that can aid future efforts to clarify the nature of the various heavy-light mesons state and their placement within the relevant spectrum.

\section{SUMMARY}\label{summary}
{{In this work, we obtained the mass  spectra of S, P, D and F wave charm-strange mesons within a relativistic framework employing a four-vector plus scalar power-law potential. Our mass predictions for well established low lying states exhibit excellent agreement with experimental measurements, thereby validating the reliability of the model. Extending beyond spectroscopy, we have applied heavy quark effective theory at leading order to systematically analyze the strong decay properties of experimentally observed excited charm-strange mesons. Using methods based on heavy quark spin and flavour symmetries, mesons can be classified in doublets. Present analysis combined with the computed mass spectra and decay widths supports the assignment of $D_{sJ}(2700)$ as \small{$2^3S_1$}, $D{s0}(2590)$ as \small{$2^1S_0$}, $D_{s3}(2860)$ as \small{$1^3D_3$}, $D_{s1}(2860)$ as \small{$1^3D_1$} and $D_{sJ}(3040)$ as \small{$2^1P_1$} states. We have presented decay width ratios that are independent of the strong coupling constants appearing in the effective Lagrangian. By systematically observing and analyzing the obtained partial decay widths, we extract the effective hadronic couplings $g_T$, $g_X$, $g_Y$, $\tilde{g}_S$ and $\tilde{g}_H$. These extracted couplings then enable us to make  predictions for the dominant decay channels and partial widths of several excited charm-strange mesons that remain to be observed. For instance, the $D^*K$ channel emerges as the most favorable mode for experimental search of \small{$D_s(1^3D_2)$} and \small{$D_s(1^1D_2)$}, whereas the $DK$ channel is favored for \small{$D_s(2^3P_0)$}.
The predicted mass spectra and decay properties along with spin-parity  assignments in present study could be valuable insights to ongoing theoretical and experimental investigations in  understanding heavy-light meson dynamics.}}

\begin{acknowledgments} 
B.P. gratefully acknowledges the support and resources provided by the Indian Institute of Technology Kanpur during the course of this work.
\end{acknowledgments}
\bibliographystyle{unsrt}   
\bibliography{ref}         
\end{document}